# A scalable gene network model of regulatory dynamics in single cells

Paul Bertin[1,2 †], Joseph D. Viviano[1,2], Alejandro Tejada-Lapuerta[3,4], Weixu Wang[3,4], Stefan Bauer[3,5,6], Fabian J. Theis[3,4,7], Yoshua Bengio[1,2]

[1] Mila, the Quebec AI Institute, Montreal, Quebec, Canada
[2] Université de Montréal, Montreal, Quebec, Canada
[3] Institute of Computational Biology, Helmholtz Munich, Munich, Germany
[4] School of Computing, Information and Technology, Technical University of Munich, Munich, Germany
[5] Helmholtz Munich, Munich, Germany
[6] Munich Center for Machine Learning (MCML), Munich, Germany
[7] TUM School of Life Sciences Weihenstephan, Technical University of Munich, Munich, Germany
[†] Correspondence to < bertinpa@mila.quebec >

## Abstract

Single-cell data provide high-dimensional measurements of the transcriptional states of cells, but extracting insights into the regulatory functions of genes, particularly identifying transcriptional mechanisms affected by biological perturbations, remains a challenge. Many perturbations induce compensatory cellular responses, making it difficult to distinguish direct from indirect effects on gene regulation. Modeling how gene regulatory functions shape the temporal dynamics of these responses is key to improving our understanding of biological perturbations. Dynamical models based on differential equations offer a principled way to capture transcriptional dynamics, but their application to single-cell data has been hindered by computational constraints, stochasticity, sparsity, and noise. Existing methods either rely on low-dimensional representations or make strong simplifying assumptions, limiting their ability to model transcriptional dynamics at scale. We introduce a Functional and Learnable model of Cell dynamicS, FLeCS, that incorporates gene network structure into coupled differential equations to model gene regulatory functions. Given (pseudo)time-series single-cell data, FLeCS accurately infers cell dynamics at scale, provides improved functional insights into transcriptional mechanisms perturbed by gene knockouts, both in myeloid differentiation and K562 Perturb-seq experiments, and simulates single-cell trajectories of A549 cells following small-molecule perturbations.

---

Single-cell data consists of high-dimensional measurements of the transcriptomic states of cells. However, extracting insights into the regulatory functions of genes from single-cell data – such as understanding which transcriptional mechanisms are affected by biological perturbations or drive differentiation – remains a challenge. Many biological perturbations are thought to directly affect specific chemical entities and cellular mechanisms. However, perturbations usually trigger a compensatory response such as immune activation[1] in order to restore homeostasis, resulting in widespread differences between control and perturbed cell states. Therefore, disentangling direct from indirect perturbational effects on cell dynamics is crucial to deepening our understanding of transcriptional mechanisms.

Transcriptional regulation is governed by interactions between genes, which form recurring patterns called network motifs. These motifs enable key biological functions, such as switching between alternative states[2], speeding up responses[3] or maintaining equilibrium[4]. Many of these motifs contain cyclic patterns that are essential for their function. Therefore, cyclic interactions between regulators and genes, along with the temporal nature of biological processes, are central to transcriptional mechanisms, and should be incorporated into computational models[5]. Dynamical models based on differential equations provide a natural framework for capturing cellular dynamics from data[6]. Given temporal observations, training such models requires computing the gradients with respect to their parameters, typically achieved via the forward sensitivity or adjoint methods[7,8], which rely on integrating augmented systems of differential equations. These computations can be computationally intensive for models with many parameters, or systems with high dimensional state spaces[9] such as the gene expression space. Additionally, sparse and noisy observations can further complicate the training process[10].

Single-cell data captures intercellular variability, but is highly noisy, and time-series observations are often sparse, leading to challenges in the application of dynamical models. Gene Regulatory Network (GRN) inference methods have been developed for single-cell data[11,12]. They typically rely on regressing regulated genes from potential regulators[13,14], therefore overlooking the temporal nature of transcriptional regulation. CellOracle[15] adopts an intermediate approach: it first learns interactions by regressing genes from their regulators, then estimates dynamics in downstream analysis by comparing the predictions of the default and intervened models. For single-cell trajectory data (*i.e.* time-series single-cell data), optimal transport can be used to match cells observed at different time points[16–18], after which dynamical models can be trained to predict the fate of individual cells over time[19–22]. Current methods typically rely on black-box models and operate in low-dimensional spaces, such as principal components or latent spaces. Some approaches incorporate gene networks with differential equations but, so far, have only been applied to low-dimensional projections of single-cell data[23] or have found a way around gradient computations using the first-order finite difference approximation[24,25] or steady state assumption[13]. Additionally, proxy information about cell dynamics – such as spliced-unspliced RNA ratios or metabolic labeling – can be used to infer dynamics[26–29]. For example, RegVelo[29] relies on a gene network and infers gene-specific dynamics from spliced-unspliced RNA ratios, rather than inferring joint dynamics from trajectory data. These approaches rely on strong assumptions about splicing rates, are prone to biases[30], and cannot be applied to datasets that lack such information.

In this work, we introduce a Functional and Learnable model of Cell dynamicS. FLeCS models cells through a set of coupled nonlinear differential equations wherein production rates are determined by a gene network – typically containing cyclic motifs – and efficiently computed via message passing[31]. Our approach learns dynamics that align with observed trajectories by integrating cellular dynamics in gene space and computing model gradients through the adjoint method, without relying on steady state assumptions or finite difference approximations. FLeCS incorporates gene network-based constraints, such as those derived from scATAC-seq, to guide the model towards biologically meaningful functions[14]. By modeling genes and their interactions without relying on low-dimensional projections, FLeCS allows for targeted modifications to the model that reflect known aspects of biological perturbations, such as loss of function of a transcript. To the best of our knowledge, FLeCS is the first method to (1) infer cell dynamics at scale from single-cell trajectory data while explicitly relying on a gene network, and (2) generate single-cell trajectories at scale without relying on a low-dimensional embedding. Our approach enhances functional insights into transcriptional mechanisms from single-cell data: it identifies gene sets that are more strongly aligned with known biological processes compared to differential expression analysis, and it learns gene interaction functions (such as up- or downregulation) and how these are affected by perturbations.

We show that FLeCS: (1) accurately infers transcriptional dynamics through a benchmark against state-of-the-art methods; (2) identifies mechanisms perturbed by gene knockouts, both in the context of myeloid differentiation and within Perturb-seq experiments on a human cell line; (3) provides functional insights consistent with known biological processes; (4) simulates single-cell trajectories over time, and (5) simulates trajectories under small-molecule perturbations.

# Results

## Scalable estimation of transcriptional dynamics via gene network models

Interactions between transcription factor genes and their regulated target genes govern transcriptional dynamics during differentiation and cellular response to biological perturbations. Given a single-cell dataset and a putative gene network (Fig. 1a), FLeCS infers key parameters such as regulatory interaction strengths (how strongly a transcription factor gene influences a target gene's production rate) and mRNA half-lives (which determine gene decay rates) (Fig. 1b). We assume that the actual regulatory interactions occurring in cells are a subset of the interactions present in this network. Experiments are conducted at scale on gene networks containing ~$10^6$ regulatory interactions among thousands of genes (4000 in our experiments). By using differential equations inspired by those describing transcriptional regulation in Systems Biology, FLeCS models, learns and integrates transcriptional dynamics in gene expression space. Optionally, the gene network can be modified to fit different experimental conditions. For instance, a connection can be removed to account for the loss of function of a transcript following a CRISPR knockout (KO). Finally, FLeCS-derived functional insights can be validated using ontology databases.

Single-cell data undergoes preprocessing before training. Since FLeCS is trained on (pseudo)time series of cells, each cell in the dataset must be assigned to a temporal group (Fig. 1c). This can be achieved via pseudotime inference methods[32,33], which associate each cell with a continuous value representing its position along its differentiation branch. This value is then quantized into discrete temporal groups. If the dataset contains cells sequenced at different acquisition times, or if cells are already associated with known differentiation stages, these timestamps can be used directly. Next, Optimal transport matching is applied to pair cells across consecutive temporal groups (Fig. 1d). Finally, these cell matchings are used to construct (pseudo)time-series, approximating how a given cell might have evolved over multiple time points (Fig. 1e).

We denote the expression levels of genes, or single-cell state, at time $t$ as a vector $G(t) = (G_1(t), ..., G_n(t)) \in \mathbb{R}^n$. In FLeCS, transcriptional dynamics are modeled through a Causal Kinetic Model (CKM), where the single-cell state $G(t)$ evolves according to a system of coupled ordinary differential equations (ODEs). The initial conditions are specified by the gene expression profiles of the earliest cells in the (pseudo)time-series, which typically correspond to progenitors in the context of differentiation. The time derivative of a gene's expression level is determined by the difference between the production rate and the decay rate of the mRNA coded by that gene (Fig. 1f). Production rates are computed based on the gene network, such that a gene's production rate at time $t$ depends on a linear function of the expression levels of its regulators at time $t$. A sigmoid activation function ensures that production rates remain positive and bounded. In this framework, the network structure is unconstrained, allowing for cyclic regulation motifs, including autoregulation. Production rates are computed efficiently at scale using message passing[31]. For each gene, the decay rate is assumed to be exponential and depends only on its own expression level. Some variants of FLeCS can include unobserved variables (Extended Data Fig. 1), see Methods for more detail.

Importantly, the functions governing production and decay rates are time-invariant, meaning they do not directly depend on time. This design choice reflects our goal to infer time-invariant regulatory rules. However, the effect of a gene $j$ on another gene $i$, $\frac{dG_i}{dG_j}$ is indirectly time-dependent due to the non-linearity in the production rate function and its dependence on the single-cell state $G(t)$. As a result, a given interaction can be active or inactive at different times, but the underlying regulatory strength remains unchanged. This strength parameter aims to capture the aggregated effect of various chemical constants including the impact of transcription factor binding on the affinity between the promoter and RNA polymerase .

During training, FLeCS jointly integrates cell dynamics to learn from single-cell trajectories. Consider a series of observed single-cell expression profiles $G(t_0)$, …, $G(t_k)$ (Fig. 1g). A differentiable ODE solver[34] forecasts $\hat{G}(t_1)$, …, $\hat{G}(t_k)$ given the initial state $G(t_0)$ by integrating the dynamics $\frac{dG}{dt}$ over time. At training time, a mean square error loss function $\text{MSE}[(G(t))_t, (\hat{G}(t))_t]$ evaluates how well predictions match observations. Gradients with respect to parameters are computed using the adjoint method, and the parameters are updated accordingly.

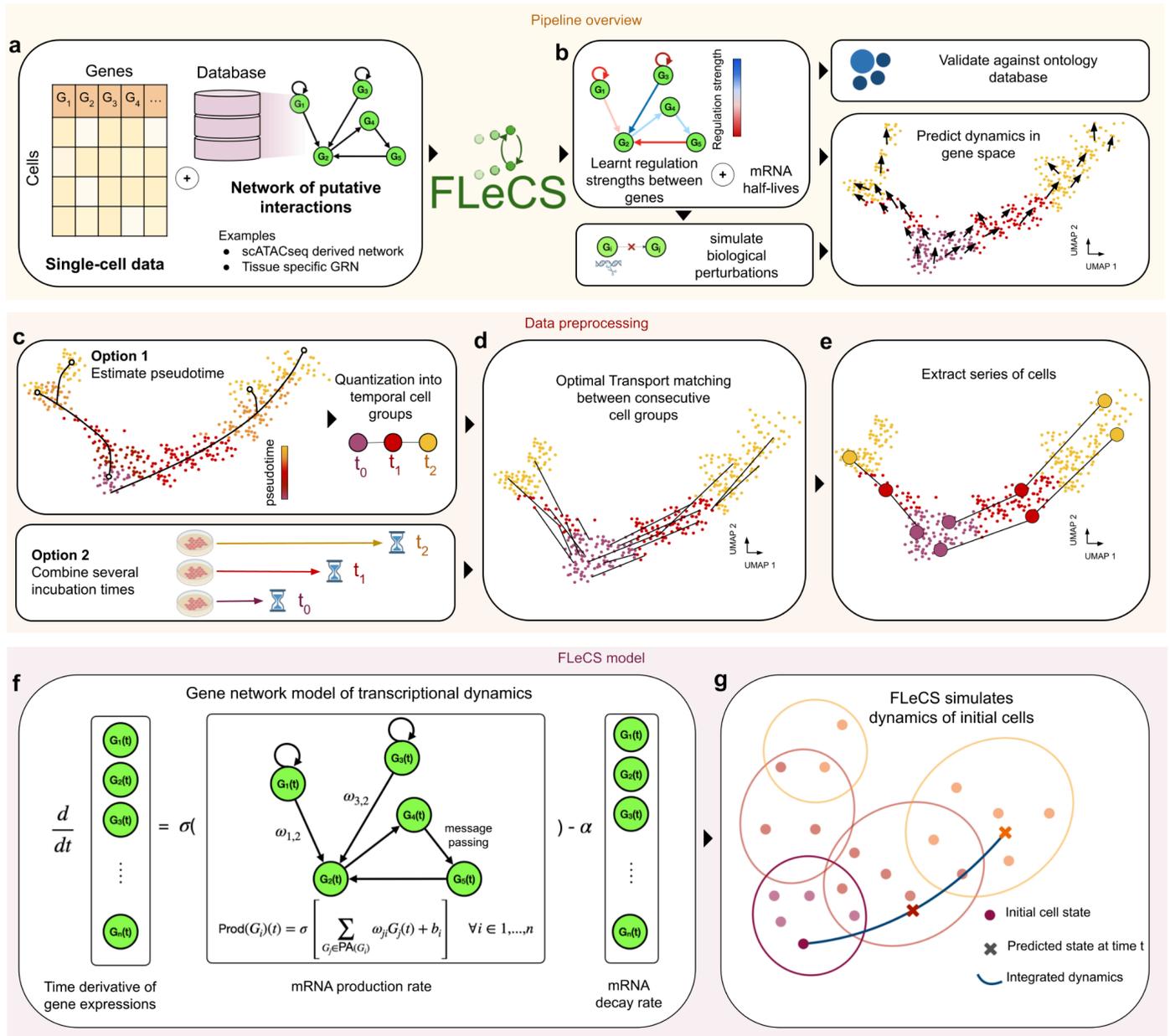

**Fig. 1 | Scalable estimation of transcriptional dynamics via gene network models. a**, Input to the pipeline is a single-cell dataset and a network of putative genetic interactions. **b**, FLeCS infers genetic interaction strengths governing mRNA production rates, as well as decay rates, which can be validated with an ontology database. Optionally, the model can be intervened on to account for biological perturbations. FLeCS then predicts cellular dynamics in gene space. **c**, During preprocessing, cells are clustered into several temporal groups, which can be derived from pseudotime if observations are made at a single time point. **d**, Cells in consecutive groups are matched through Optimal Transport. **e**, These matchings are used to extract time series of individual cells on which FLeCS is trained. **f**, Cells are modeled as a system of coupled ordinary differential equations. mRNA production rates are based on a gene network and computed efficiently at scale using message passing. mRNA decay rates are exponential. **g**, During training FLeCS simulates the dynamics of initial cells, and learns from differences in predicted cell states and observed ones at later time points.

# FLeCS accurately infers transcriptional dynamics

We first demonstrate FLeCS's ability to accurately infer transcriptional dynamics and evaluate its performance against state-of-the-art methods, using a single-cell dataset of myeloid progenitor development[35]. Cells are visualized using uniform manifold approximation and projection (UMAP) (Fig. 2a). The dataset contains myeloid cells, sorted common myeloid progenitor (CMP) cells, and two batches of cells perturbed by knockouts of the Cebpa and Cebpe genes (Fig. 2b). During preprocessing, optimal transport is used to match progenitors to unsorted myeloid cells. Then, pseudotime-series of cells of length 10 are constructed (Fig. 2c) by (1) identifying the shortest path in the cell-to-cell k-NN graph between a progenitor cell and its matched unsorted myeloid cell, and (2) subsampling cells along that path. Details on the construction of pseudotime-series of cells are provided in Methods, with comparisons between alternative approaches in Extended Data Fig. 2.

FLeCS is provided with a putative gene regulatory network among $n$ genes inferred from single-cell chromatin accessibility (scATAC-seq) data[36]. We assume that the actual regulatory interactions occurring in cells are a subset of the interactions $E$ present in this network (Fig. 2d). All experiments are performed using $n = 4,000$ most highly variable genes. After training, the dynamics $\frac{dG}{dt} \in \mathbb{R}^n$ predicted by FLeCS are embedded into the two-dimensional UMAP space for visualization (Fig. 2e). We provide similar visualizations (Extended Data Fig. 2e) for (1) variants of FLeCS that incorporate latent variables or unobserved protein proxy variables, (2) state-of-the-art cell dynamics inference methods including OT-CFM[37] and scTour[22], and (3) a first-order heuristic baseline, in which cell dynamics simply point towards the next cell in the pseudotime-series extracted during preprocessing. FLeCS and its variants infer smooth dynamics from progenitors towards differentiated myeloid cells, aligning well with observed differentiation branches.

We further evaluate these cell dynamics inference methods on held-out test cells, with smoothness of dynamics confirmed using velocity confidence (VC), which quantifies the consistency of predicted dynamics among neighboring cells. FLeCS and its variants achieve the best performance (Fig. 2f). In particular, FLeCS significantly outperforms the heuristic baseline that follows the extracted cell trajectories used for training, highlighting its denoising capabilities. Additionally, the distribution of regulation strengths learned by FLeCS is highly sparse and concentrated around zero (Fig. 2g), suggesting that only a small proportion of the putative interactions in the scATAC-seq-derived network are active in myeloid differentiation.

Finally, we demonstrate how FLeCS can be adapted to interventional settings. The model was retrained on data from either the Cebpa or Cebpe knockout, using a modified network in which all outgoing connections of the knocked-out gene are removed to reflect the assumption that its transcript has undergone a complete loss of function (Fig. 2h). For each knockout, we extracted cell trajectories from progenitor cells to the corresponding knockout population (Fig. 2i), and trained a FLeCS model. Predicted cell dynamics from each model are projected into UMAP space (Fig. 2j). In the Cebpa knockout condition, FLeCs dynamics indicate that all progenitor cells differentiate towards the megakaryocyte–erythroid lineage (left branch), which is consistent with the absence of granulocyte-monocyte lineage cells (right branch) in the Cebpa knockout population. Conversely, in the Cebpe knockout condition, FLeCS dynamics indicate that most progenitor cells differentiate towards the granulocyte-monocyte lineage.

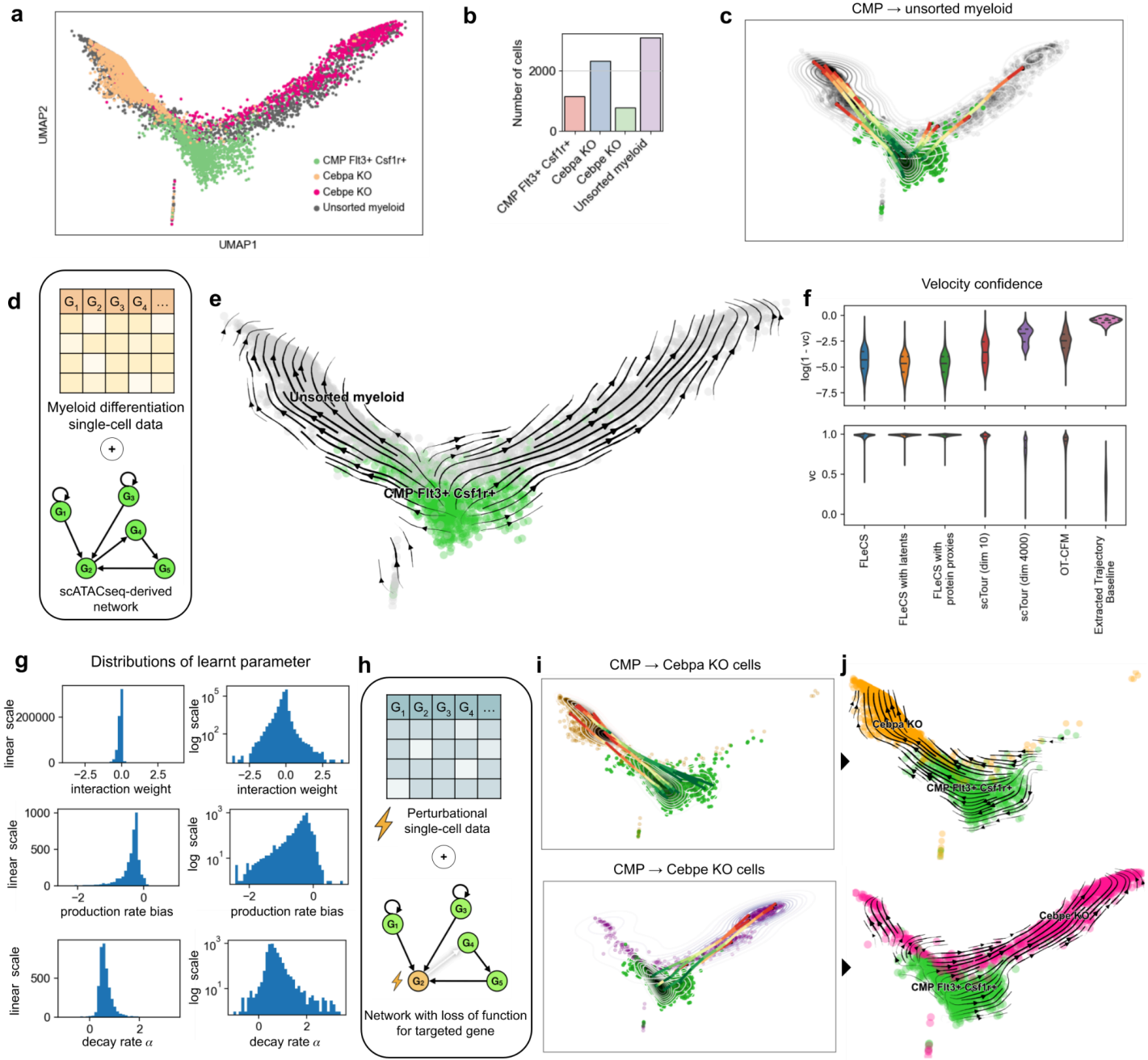

**Fig. 2 | FLeCS accurately infers transcriptional dynamics of myeloid differentiation. a**, UMAP representation of the myeloid progenitor development dataset, including sorted progenitors, and cells affected by two distinct CRISPR knockouts. **b**, Number of cells within each subpopulation. **c**, Sample of the trajectories extracted during preprocessing, mapping progenitor cells (green) to the unsorted myeloid population (grey). **d**, FLeCS is first trained on trajectories extracted from the myeloid differentiation data, using a network of putative interactions derived from scATAC-seq data. **e**, Projection into UMAP space of the dynamics learnt by FLeCS in gene space. **f**, Distribution of velocity confidences (vc) for the dynamics learnt by FLeCS, its variants that include unobserved variables, and baselines. Computed on held out test cells. **g**, Distributions of the learnt parameters, with linear (left) and logarithmic (right) scaled y-axis. **h**, Two additional

FLeCS models are trained on the perturbational subsets of the data corresponding to knockouts on the Cebpa and Cebpe genes. The network of putative interactions is ablated to account for a complete loss of function of the perturbed gene's transcript. **i**, Sample of the extracted trajectories, mapping progenitors to Cebpa (top) and Cebpe (bottom) knocked out cells. **j**, Projections into the UMAP space of the dynamics learnt from perturbational data.

## FLeCS identifies mechanisms perturbed by gene knockouts

We now investigate in more detail the insights that can be derived from FLeCS. It is important to note that there is not a unique set of parameters (*i.e.*, a single trained model) consistent with the observed data; multiple functional explanations can lead to the same measured outcome (in our case, dynamics consistent with observed trajectories). Therefore, to obtain reliable insights, we estimate the posterior distribution $P(\omega|\mathcal{D})$ of interaction strengths $\omega \in \mathbb{R}^{|E|}$ that are consistent with the data $\mathcal{D}$, rather than relying on a point estimate.

Samples from the posterior distribution $P(\omega|\mathcal{D})$ can be obtained by training a deep ensemble[38] of FLeCS models. Ensembles of size 10 were used in our experiments. For a given interaction $G_i \to G_j$, one can focus on the marginal distribution $P(\omega_{i,j}|\mathcal{D})$. The spread of the distribution provides an estimate of the uncertainty about the true value of $\omega_{i,j}$. Given data corresponding to wild-type cells ($\mathcal{D}_{WT}$) and perturbed cells ($\mathcal{D}_{Pert}$), one can analyse the difference between the marginals $P(\omega_{i,j}|\mathcal{D}_{WT})$ and $P(\omega_{i,j}|\mathcal{D}_{Pert})$, obtained after training on $\mathcal{D}_{WT}$ and $\mathcal{D}_{Pert}$, respectively. If the two marginal distributions do not overlap, it suggests that the interaction $G_i \to G_j$ has been affected by the biological perturbation (Fig. 3a).

We performed this analysis using the previously described myeloid differentiation dataset, encompassing three biological contexts: wild-type (WT), Cebpa knockout, and Cebpe knockout. To confidently identify whether an interaction has been affected by a knockout, we conducted Welch's t-tests to compare the means of $P(\omega_{i,j}|\mathcal{D}_{Pert})$ and $P(\omega_{i,j}|\mathcal{D}_{WT})$, for all interactions in the network, across both knockouts. Given the large number of tests performed, and to limit the number of false positives, we applied a stringent significance threshold (p-value $\leq \frac{1}{|E|}$), ensuring that, in expectation, only one interaction would be falsely identified as affected. Among $\sim 6 \times 10^6$ interactions present in the gene network, we confidently identify 1,494 interactions affected by the Cebpa knockout, and 3,176 interactions by the Cebpe knockout (Fig. 3b).

Changes in regulation strengths are visualized using a scatter plot, where each interaction is positioned based on its regulation strength change after the Cebpa knockout (x-axis) and the Cebpe knockout (y-axis) (Fig. 3c). We define the change in regulation strength $\omega_{i,j}$ after a perturbation as $\mathbb{E}[\omega_{i,j}|\mathcal{D}_{Pert}] - \mathbb{E}[\omega_{i,j}|\mathcal{D}_{WT}]$. Overall uncertainty for each interaction is computed as the geometric mean of the Welch's t-tests' uncertainties $U(Cebpa\ KO,\ WT)$ and $U(Cebpe\ KO,\ WT)$; see Methods for more detail on how uncertainties are computed. The vast majority of interactions cluster around $(0, 0)$, indicating that they are not affected by either knockout, while only a few interactions deviate significantly, confirming the extreme sparsity of functional effects. We report the percentage of interactions that are (1) non-zero with high confidence for each condition, (2) affected by the perturbation for each knockout, and (3) that change sign after perturbation (Fig. 3d).

We now focus on the subnetwork most affected by the knockouts. To identify this subnetwork, we ranked interactions based on the magnitude of their impact from the two knockouts and their associated uncertainties. We then selected the first $m$ genes ($m = 33$ in our experiments) to appear within these interactions. Next, we constructed the subnetwork comprising these $m$ genes, and include all interactions among them that are non-zero in at least one condition. The posterior distribution of regulation strengths in this subnetwork is presented for each of the three conditions (Fig. 3e, Extended Data Fig. 3b). In some cases, the posteriors remain highly consistent across conditions (*e.g.* GATA2→MMP8), while in others, the posterior mean (*e.g.* CUX1→ELANE) or the uncertainty (*e.g.* TAL1→GSTM1) varies substantially across conditions. The learned interaction strengths and their associated uncertainty are further visualized within the subnetwork (Fig. 3f).

Notably, the regulatory mechanisms of PRTN3, a gene implicated in myeloid differentiation[39], appear to be particularly affected by the Cebpa knockout. Specifically, LYL1, NFYA and TCF4 – genes involved in proliferation[40,41] and development[42] within myeloid and erythroid cells – upregulate PRTN3 following the Cebpa knockout. While PRTN3 is primarily a marker of monocytes, it is also sporadically expressed in the megakaryocyte–erythroid lineage (Extended Data Fig. 3h). LYL1, NFYA and TCF4 are also expressed in this lineage (but not in the granulocyte-monocyte lineage) and upregulation relationships towards PRTN3 have been inferred by FLeCS under the Cebpa knockout condition. In contrast, the regulation of PRTN3 by STAT1 (which inhibits myeloid-derived suppressor cell accumulation[43]), GATA2 (a crucial regulator in myeloid[44]) and SMARCC1 (associated with immune infiltration[45]) remains unchanged according to FLeCS.

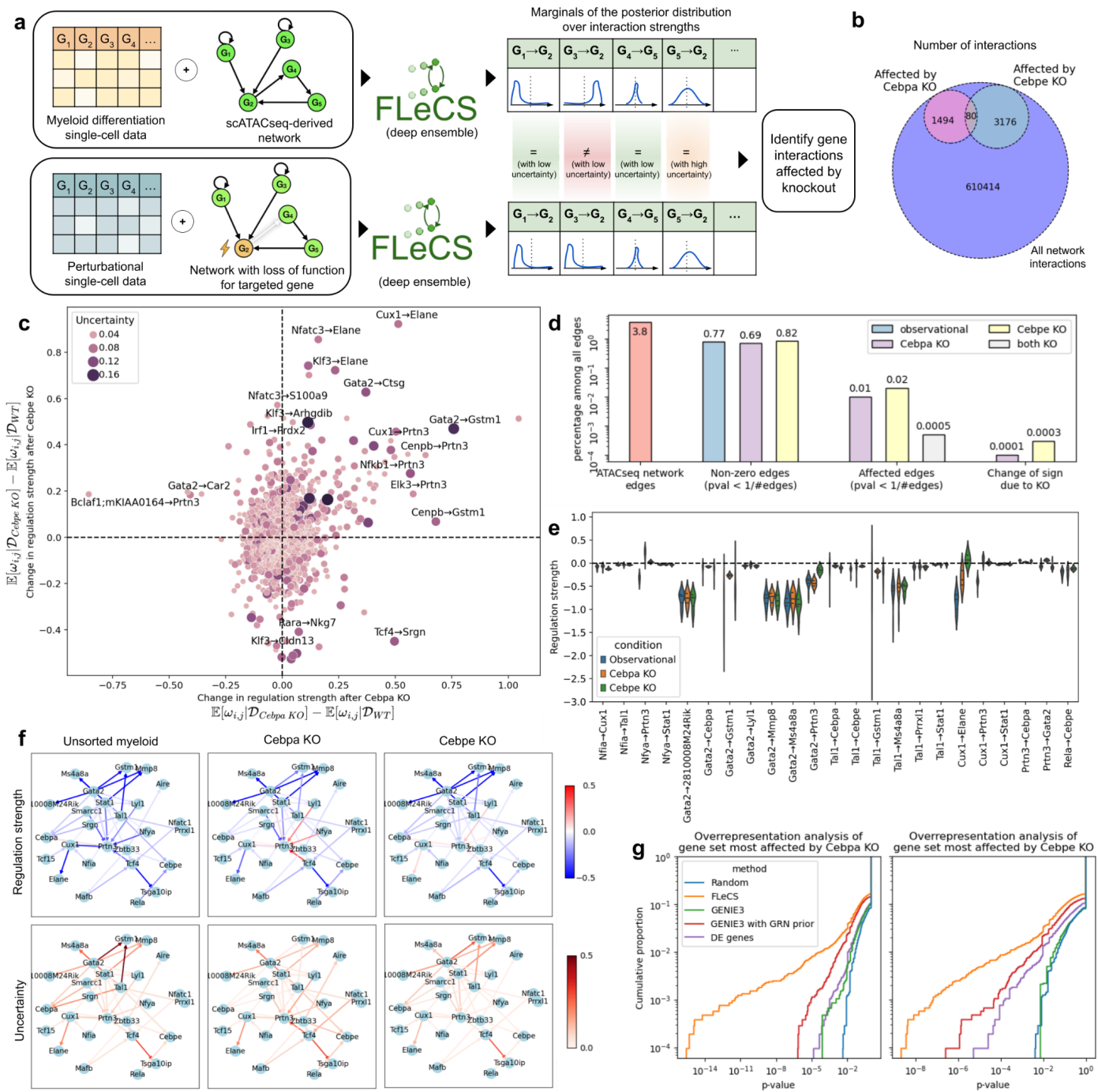

**Fig. 3 | FLeCS identifies mechanisms perturbed by gene knockouts and provides functional insight aligned with known biological processes. a**, For each condition (wild type, Cebpa KO, Cebpe KO), an ensemble of models is trained. This gives access to samples from the posterior distribution of genetic interaction strengths. Differences between the posterior distribution marginals across conditions can identify interactions that have been affected. **b**, Out of 610,414 putative interactions, we identify with high confidence (p-value ≤ 1 / #interactions) 1294 interactions that have been affected by the Cebpa knockout, and 3176 interactions by the Cebpe knockout. **c**, Scatter plot of interactions showing how they are affected by the Cebpa knockout (x-axis) and the Cebpe knockout (y-axis). Size and color represents uncertainty, which is derived from the standard deviations of the posterior distribution marginals. **d**,

Histogram of various edge proportions among all possible interactions in the complete graph. **e**, Regulation strength posteriors across the three conditions, for some edges within the most affected subnetwork. **f**, Subnetwork most affected by both knockouts. Edge colors represent the mean of the posterior (top) and uncertainty (bottom), across three conditions. **g**, Overrepresentation of the most affected subnetwork gene set across all biological processes in the PantherDB knowledgebase, for the Cebpa (left) and Cebpe (right) knockouts. Cumulative proportion of p-values from Fisher's exact tests. Results for gene sets identified via various methods, including random selection. All sets have the same size. DE: differential expression.

## FLeCS offers improved functional insights

We then aimed to assess whether FLeCS's insights are consistent with known biological processes. To validate its functional insights, we performed an overrepresentation analysis using the PantherDB knowledgebase[46]: we assessed the enrichment of all Gene Ontology[47] (GO) biological process terms within the set of genes in the most affected subnetwork, using the 4000 highly variable genes modeled by FLeCS as the reference set.

We compared FLeCS against standard differential gene expression analysis, the GRN inference method GENIE3[13], as well as a variant in which the interactions inferred by GENIE3 are restricted to those present in the scATAC-seq-derived network used to constrain FLeCS. For each method, the same most-affected-subnetwork extraction procedure was utilized to retrieve a set of $m = 33$ genes, which is used for overrepresentation analysis. The most enriched ontology terms according to FLeCS include the regulation of erythrocyte differentiation, definitive hemopoiesis and myeloid differentiation (Extended Data Fig. 3g). Fig. 3g displays the cumulative distributions of Fisher's exact test p-values for the different methods. Higher proportions of small p-values indicate stronger alignment with a specific part of the hierarchy of known biological processes. FLeCS outperforms other methods, demonstrating that its insights align more closely with known biology. Interestingly, constraining GENIE3 with the scATAC-seq-derived network improves its performance but still remains inferior to that of FLeCS.

We repeated this analysis on a Perturb-seq dataset of K562 cells, a cell line derived from a myelogenous leukemia patient. To be consistent with the biological context of the dataset, the interactions learned by FLeCS were constrained using a myeloid leukemia GRN from the FANTOM5 database[48] (Fig. 4a). Although the dataset includes a large number of single-gene knockdowns, most affect only a small number of cells (Fig. 4b). This poses a challenge, as our approach involves training models independently for each condition. To address this, we restricted our analysis to the 12 knockdowns that affected at least 600 cells. We computed a UMAP representation of the data and associated cells with the different phases of the cell cycle – Gap 1 (G1), Synthesis (S) and Gap2/Mitosis (G2M) – based on known marker genes (Fig. 4c). As expected for a proliferative cell line, the cell cycle is clearly distinguishable and dominates the primary axes of variability in the UMAP representation. Since knocked-down cells also undergo replication (Extended Data Fig. 4a), the outcomes after knockdowns are confounded by cells' phase in the cell cycle, complicating the analysis of knockdowns' direct effects.

For each of the 12 knockdowns and for the control cells, we extracted three sets of trajectories corresponding to transitions between successive cell cycle phases (Fig. 4d), employing the same trajectory extraction method as previously described. Subsequently, for each knockdown and the control, an ensemble of FLeCS models was trained on the combined data from these three sets of trajectories. The dynamics learned by FLeCS accurately recapitulate the cell cycle (Fig. 4e), illustrating FLeCS's capability to model periodic dynamics. This is in contrast with methods based on potential energy landscapes[17,18] – known as Waddington landscapes[49] – which are unable to capture periodicity[50].

For each knockdown condition, we identified and visualized the most affected subnetwork (Fig. 4f, Extended Data Fig. 4b). Subsequent overrepresentation analysis revealed that FLeCS's insights align more closely with known biology compared to other methods, outperforming them in 11 out of the 12 knockdowns, often by a large margin (Fig. 4g). Notably, the performance of standard differential expression analysis is nearly equal to random chance, highlighting the challenge of extracting functional insights from Perturb-seq cell line data.

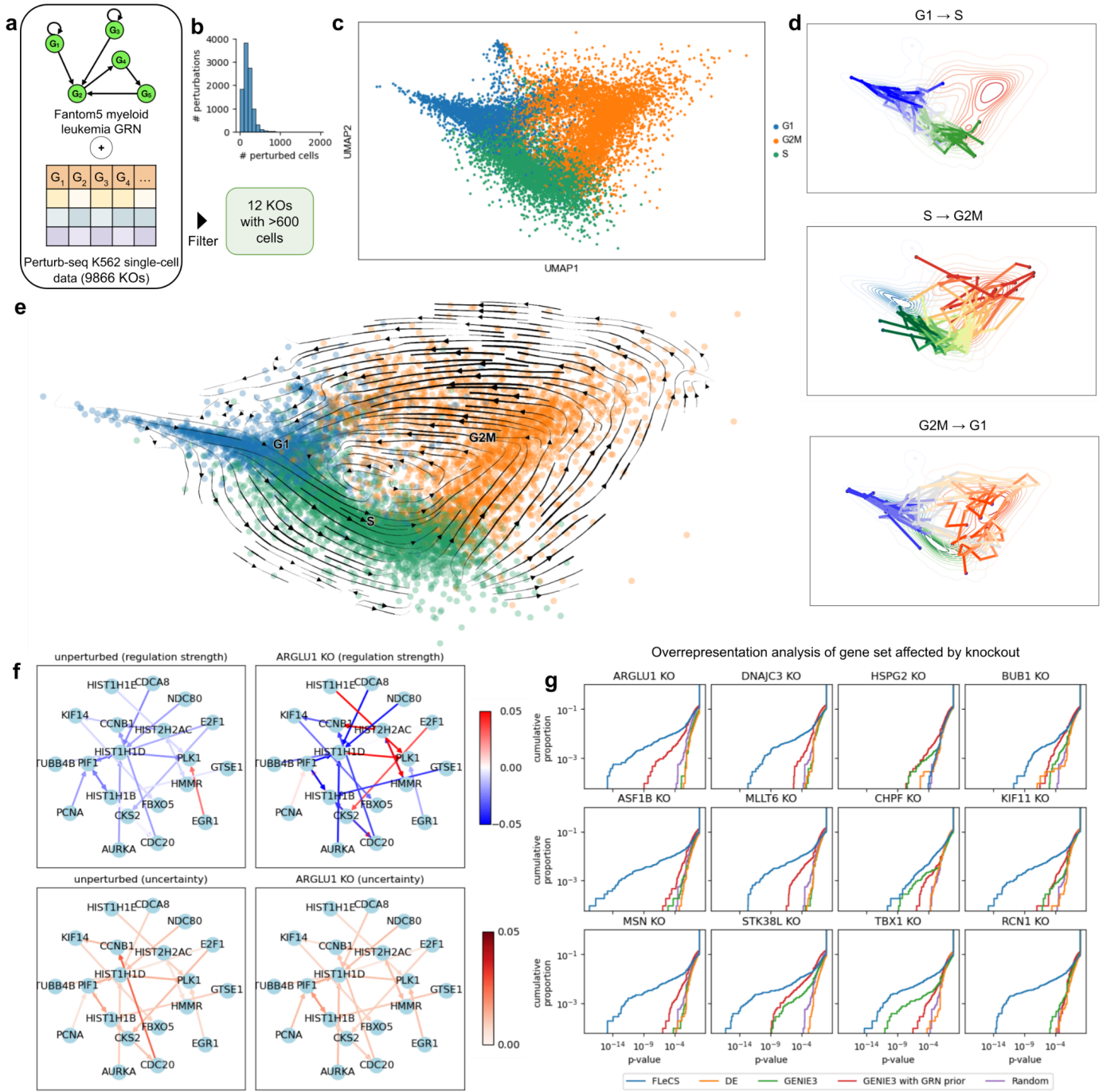

**Fig. 4 | FLeCS provides functional insight aligned with known biological processes from Perturb-seq data. a**, FLeCS is trained on Perturb-seq data on the K562 human cell line, using a GRN of myeloid leukemia from the FANTOM5 database. Knockdowns with fewer than 600 cells are filtered out. **b**, Distribution of the number of cells per knockdown, across all perturbations in the dataset. **c**, UMAP representation of the Perturb-seq dataset. Color corresponds to the inferred phase of cells in the cell cycle. **d**, Some of the trajectories extracted during preprocessing, mapping cells in one phase of the cell cycle to the next, G1→S→G2M. FLeCS is trained on the union of these three sets of trajectories. **e**, Projection into UMAP space of the dynamics learnt by FLeCS in gene space, for control cells. **f**, Subnetwork most affected by the ARGLU1 knockdown. Edge colors represent the mean of the posterior (top) and

uncertainty (bottom), for unperturbed (left) and perturbed (right) cells. **g**, Overrepresentation of the gene set of the most affected subnetwork across all biological processes in the PantherDB knowledgebase, for all knockdowns that passed filtering. Cumulative proportion of p-values from Fisher's exact tests. Results for gene sets identified via various methods, including random selection. All sets have the same size. DE: differential expression.

## FLeCS simulates single-cell trajectories over time

We now demonstrate how FLeCS can simulate cell trajectories over time and generate synthetic cells at future time points by integrating the learned dynamics. Additionally, we show that FLeCS can be adapted to capture small molecule perturbations and simulate trajectories of perturbed cells. We experiment on a single-cell dataset from A549 cells, which includes multiple chemical perturbations and two distinct incubation times[51] (Extended Data Fig5a-b). To match this new biological context, FLeCS was constrained based on a adenocarcinoma GRN from the FANTOM5 database[48] (Fig. 5a).

We first focus on control cells. The two incubation times were used to build trajectories, considering control cells incubated for 24 hours as initial cells, and those incubated for 72 hours as target cells. A FLeCS model was trained on these extracted trajectories. To improve convergence during training, we augmented the extracted trajectories by sampling intermediate cells using a procedure similar to Flow-Matching[37] (FM) (see Methods for more details). The learned dynamics were projected into UMAP space (Fig. 5b).

We then integrated the learned cell dynamics and generated a trajectory of 10 cells for each initial control cell. We confirmed that the generated trajectories converge towards the population of control cells observed at 72 hours, as measured by the Earth Mover's distance (Fig. 5c). We notice that FM-style data augmentation improves stability of generated trajectories, especially at later time points. A new UMAP representation was computed for the synthetic dataset, with a few synthetic trajectories highlighted (Fig. 5d).

## FLeCS simulates trajectories under small molecule perturbations

We now adapt FLeCS to the context of small molecule perturbation. Structural information about the small molecules is provided as Morgan fingerprints, and FLeCS can be modified – *i.e.*, intervened on – to generate trajectories for cells perturbed by a given molecule (Fig. 5e). Given a drug fingerprint $v$, an interventional model predicts the effects of the drug on production rates $E_p(v) \in \mathbb{R}^n$ and on decay rates $E_d(v) \in \mathbb{R}^n$, which are then used to offset the default production and decay rates, respectively (Fig. 5f). To focus on drugs with a strong perturbation effect, we trained a binary tree classifier for each drug to distinguish drug-perturbed cells from control cells. Only drugs with test accuracy exceeding $0.8$ on held-out test cells were included in our analysis (Extended Data Fig. 5c). The interventional model operates on the default FLeCS model that captures the dynamics of control cells. During this training stage, the parameters of the default FLeCS model were fixed, and only the interventional model predicting $E_p(v)$ and $E_d(v)$ was trained on drug-perturbed cells.

We generated trajectories of perturbed cells for drugs included in the training set. As examples, we visualize generated trajectories for Belinostat and Abexinostat using Principal Component Analysis (PCA) (Fig. 5g, Extended Data Fig. 5d). We confirmed that generated trajectories converge towards the target population of perturbed cells: for each drug that passed filtering with accuracy $> 0.85$, we computed the Earth Mover's distance (EMD) to held-out test cells from the target population. We evaluate trajectories generated by the intervened FLeCS model that captures the effect of the drug (Fig. 5h, Extended Data Fig. 5e), and trajectories generated by the default FLeCS model without any drug perturbation (referred to as observational). We benchmarked against scGen[52], a Variational Autoencoder based model that does not take temporal dynamics into account. To contextualize these results, we computed a lower bound on EMD by randomly splitting the target population into two halves and measuring the distance between them. We observe that FleCS performs similarly to, though slightly worse than, scGen. This can likely be attributed to the challenge of training a functional model at scale without relying on a low-dimensional latent space. For completeness, we also report performance on held-out test drugs that were not seen during training (Extended Data Fig. 5f). Performance varies widely across drugs, with some cases of catastrophic failure (*e.g.* Triamcinolone Acetonide). This suggests – unsurprisingly – that the limited set of drugs used for training does not offer a representative view of the full space of possible drug structures, limiting generalization to new drugs.

Genes whose production rate and decay rate are most affected under different drug perturbations are visualized (Fig. 5g, Extended Data Fig. 5g). We selected the top $m$ genes ($m = 33$ in our experiments) whose decay rates are most impacted by the drug, and performed an overrepresentation analysis using the PantherDB database[46]. Except for a slight improvement of FLeCS over differential gene expression analysis in the case of Abexinostat, both methods perform close to the level of randomness for other drugs (Fig. 5h, Extended Data Fig. 5h). This analysis indicates that, while our approach recapitulates cell trajectories under small molecule perturbations, the inferred direct effects of drugs do not align with known biological mechanisms. This discrepancy could be due to the complexity of cellular responses to small molecule perturbations, which may directly impact many more mechanisms compared to targeted CRISPR perturbations, or to limitations in how drug interventions have been modeled within FLeCS.

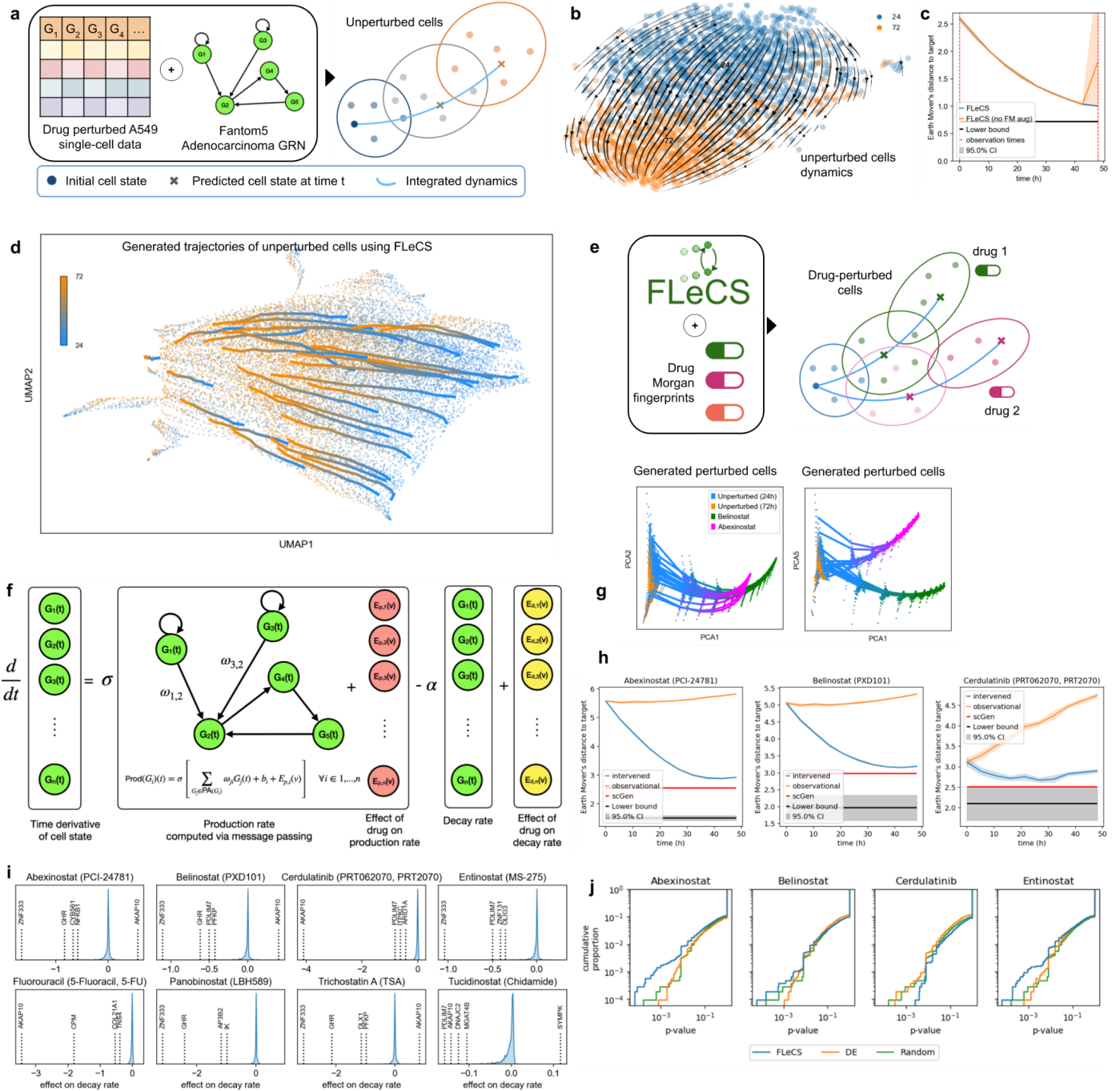

**Fig. 5 | FLeCS captures small molecule perturbations and simulations of transcriptional dynamics over time. a**, FLeCS is trained on a drug screen single-cell dataset on the A549 human cell line, using an adenocarcinoma GRN from the FANTOM5 database. Cell trajectories are generated by integrating the learnt dynamics. **b**, Projection into UMAP space of the dynamics learnt by FLeCS in gene space, for control cells. **c**, Earth Mover's distance between generated cells and held out test cells from the target population of control cells, as a function of time along generated trajectories. Results for FLeCS trained with and without Flow-Matching (FM)-style data-augmentation. A lower bound on achievable performance is shown in grey. CI: confidence interval. **d**, UMAP representation of synthetic cells generated by FLeCS under the control condition. Some generated trajectories are highlighted. **e**, FLeCS can generate

trajectories of perturbed cells when expanded with an interventional model taking Morgan fingerprints as drug features, **f**, Drug effects on production and decay rates are modeled through two MLPs that take Morgan fingerprints as input. **g**, Principal component representation of synthetic cells generated by FLeCS for the control condition and under two drug perturbations, Abexinostat and Belinostat. Lines correspond to generated trajectories. **h**, Earth Mover distance between generated cells and held out test cells from the target population of perturbed cells, for different drugs included in the training set. X-axis corresponds to time along generated trajectories. For each drug, results for FLeCS in the control setting (observational), and when FLeCS is intervened on to capture the drug effect. scGen does not take time into account, therefore performance is represented as a horizontal line. A lower bound on achievable performance is shown in grey. CI: confidence interval. **i**, Distribution of drug effects on decay rates across all genes, for 8 different drugs. Most affected genes are highlighted. **j**, Overrepresentation of the set of most affected genes across all biological processes in the PantherDB knowledgebase. All sets have the same size. DE: differential expression.

# Discussion

In this work, we introduce FLeCS, a scalable gene network model of single-cell dynamics that is rooted in the models of transcriptional regulation developed in Systems biology. Gene transcriptional dynamics are governed by a production rate dependent on the concentration of the gene's regulators, and an exponential decay rate. FLeCS leverages a gene network of putative interactions that can be chosen based on the biological context, and allows for cyclic regulatory motifs. It scales to thousands of genes and hundreds of thousands of interactions. Additionally, it can incorporate various biological perturbations, such as CRISPR knockouts or small molecule perturbations. FLeCS learns regulation strengths and decay rates from single-cell data. (Pseudo)time-series of cells are first extracted from the data, then used to train a system of coupled differential equations. The model infers and predicts cell dynamics in gene space and simulates single-cell trajectories under different conditions. Crucially, FLeCS does not rely on low-dimensional embeddings which often reduce model interpretability. It also avoids common approximations – such as first-order finite differences – and simplifying assumptions, including steady-state conditions or the existence of a potential energy landscape, which are not relevant in all biological contexts.

We have applied and evaluated FLeCS across different biological contexts, namely mouse myeloid differentiation, CRISPR knockdowns in a human myeloid leukemia cell line, and small molecule perturbations in a human adenocarcinoma cell line. Our results demonstrate that FLeCS accurately infers transcriptional dynamics (Fig. 2), identifies mechanisms affected by gene knockouts and knockdowns, and provides improved functional insights compared to state-of-the-art methods (Fig. 3-4). Additionally, FLeCS simulates single-cell dynamics over time under both control and drug-perturbed conditions (Fig. 5).

We now discuss the limitations of our approach. FLeCS currently relies on production rates that are based on a linear combination of the expression levels of the regulators. However, non-additive effects between regulators[53] exist. Exploring broader families of functions that include interaction terms between the different regulators, such as multidimensional generalizations of Hill equations[54] or multilayer perceptrons (MLPs) is an exciting avenue for future work. This approach would complicate model training and involve increased computational costs.

FLeCS currently relies on deterministic differential equations, but biological processes are inherently stochastic. Given that FLeCS already utilizes separate production and decay rates, extending it to generate at test time stochastic trajectories using the Gillespie algorithm[55] to simulate a jump process would be a natural progression. However, modeling stochastic trajectories at training time will require additional effort, as current differentiable stochastic differential equation solvers[56] only handle Wiener processes, which do not align with the discrete nature of biochemical reactions involving count variables.

The current pipeline relies on the extraction of (pseudo)time-series of cells which are then used to train the production and decay rate parameters. However, in some cases, this extraction procedure may fail (*e.g.* with series involving jumps across distinct differentiation branches), impacting the biological relevance of the FLeCS model trained downstream. Further refinement of this extraction procedure is needed, potentially leveraging advanced optimal transport methods that account for cell density in expression space and imbalances between cell populations at different time points[19,57]. Moreover, instead of deterministically extracting trajectories ahead of time, one could opt for an extraction approach based on cell transition probabilities[58] in which trajectories are sampled dynamically.

FLeCS currently relies on hard constraints provided by the structure of the gene network. We have demonstrated that the approach effectively discards most putative interactions by learning near-zero strengths for the vast majority of them. Future work could explore strategies to iteratively refine the gene network structure by applying hard constraints on interactions found to have near-zero strengths, and possibly adding new interactions.

Training FLeCS at scale using the adjoint method can be challenging. Denoising the data through MAGIC[33], and relying on bounded production rates to control the regularity of dynamics, as well as the associated computational and memory costs necessary for integration[9], were key to successfully training models. Further efforts should be made to simplify and enhance the robustness of the training process.

Currently, distinct parameters are learned for each perturbational condition. For small molecule perturbations, the interventional model predicting drug effects on production and decay rates is trained across conditions, once production and decay rate parameters have been learned from control cells. In contrast, for CRISPR knockouts, a separate model is trained for each condition. Ideally, one would jointly learn a unique set of parameters capable of recapitulating all conditions. The current approach assumes that knockouts result in a loss of function of the targeted gene. However, we observed that this targeted change alone – the removal of edges outgoing the targeted gene – did not provide enough flexibility to fit vastly different dynamics across conditions when all other interactions remain unchanged. Training models across multiple conditions is a challenging optimization problem and an active area of research in the machine learning community[59,60]. Adapting such methods to FLeCS presents an exciting direction for future work.

Most of our experiments involve a model that only accounts for gene variables capturing expression levels. However, it is known that many other factors influence cell fate, including protein levels (for which mRNA levels are only a proxy), the 3D structure of the cell –including variations in concentrations across subcellular compartments–, metabolite levels, and environmental signals like cell-cell communication. FLeCS is an extensible framework for modeling cellular dynamics at a functional level. We have presented two variants that model protein proxy variables and latent variables (Extended Data Fig. 1), and evaluated the learned dynamics (Fig. 2g). These variants would be particularly appropriate when applied to richer data modalities, for instance, where protein levels are measured alongside mRNA levels. More generally, approaches that can jointly learn from transcriptomics, proteomics, and metabolomics should be investigated.

FLeCS has so far been applied to relatively small datasets, limited either by the number of conditions, or the number of cells per condition. Consequently, we have not yet demonstrated generalization to novel conditions. The development of large interventional cell atlases is being envisioned[61] and could provide a foundation for training more extensive versions of FLeCS. Investigating different ways to model interventions within these large datasets, and exploring whether generalization to novel perturbations can be achieved, is a particularly exciting direction for future research. Such generalization capabilities would allow models to be integrated within an active learning pipeline[62–64] to suggest informative or promising perturbations to be validated experimentally.

Gene network models of cell dynamics are a valuable tool for improving our functional understanding of biological perturbations. FLeCS is a first step towards improved functional insights into transcription mechanisms from single-cell data.

---

# Methods

## Theoretical background

**Causal Kinetic Models.** Causal Kinetic Models[65] (CKMs) are a class of causal models for dynamical systems. Consider a system with state variables at time $t$ $(x_1(t), ..., x_n(t)) \in \mathbb{S}$, where $\mathbb{S}$ is typically $\mathbb{R}^n$ or $\mathbb{R}_+^n$ in the case of gene expressions. In their deterministic version, CKMs consist in a system of coupled ordinary differential equations (ODEs) and initial conditions $\xi$:

$$\frac{d}{dt}x_i(t) := f_i(\text{PA}(x_i)(t)), \qquad x_i(0) := \xi_i, \qquad \forall i \in 1, ..., n$$

where $\text{PA}(x_i) \subseteq \{x_1, ..., x_n\}$ is the set of parent variables of $x_i$ that influence its dynamics, and $\text{PA}(x_i)(t)$ refers to their values at time t. Therefore, in CKMs, the evolution of a given variable at time $t$ depends on the state of other variables in the system – its parents – at time $t$.

Let us consider the graph $\mathcal{G}$ with nodes $(x_1, ..., x_n)$ and edges going from parents $\text{PA}(x_i)$ to $x_i$ for $i \in 1, ..., n$. No constraint is imposed on $\mathcal{G}$, in particular, it can contain cycles and autoregulation patterns (i.e. $x_i \in \text{PA}(x_i)$). Note that the functions $f_i$ do not directly depend on time, therefore defining time-invariant rules that govern the evolution of the system.

The state of the system at time $t > 0$ can be computed by jointly integrating the dynamics of variables $x_1, ..., x_n$:

$$x_i(t) = \xi_i + \int_0^t f_i(\text{PA}(x_i)(s))ds \qquad \forall i \in 1,..., n$$

One or several differential equations in the system can be modified through so-called interventions. These modifications typically consist in inserting and deleting edges in the graph $\mathcal{G}$, or altering the functions $f_i$ (e.g. modifying their parameters).

**Neural Ordinary Differential Equations and adjoint method.** Neural Ordinary differential Equations[34] (neural ODE) are a class of machine learning models that rely on a system of ordinary differential equations (ODEs) $\frac{d}{dt}x(t) := f_\theta(x(t), t)$, where $x(t) \in \mathbb{R}^n$, as well as some initial conditions $\xi$. $f_\theta$ is a learnable function with parameters $\theta \in \Theta$, typically a multilayer perceptron (MLP). The state of the system $x(t)$ at time $t > 0$ can be computed by integrating the dynamics forward in time using an ODE solver.

The model is trained via the adjoint method[8] which computes gradients with respect to $\theta$ by solving an augmented ODE backward in time (satisfied by the adjoint variable of the Lagrangian, from which the method derives its name), significantly reducing computational and memory costs compared to backpropagating through the operations of the ODE solver. However, the adjoint method is subject to numerical errors or instabilities that can be introduced when solving the system backward.

**Message passing.** Message passing refers to propagation and aggregation of information among the nodes of a graph. Typically, at each step, a node $x_i$ receives some information $\mathrm{agg}(\{m_{j \to i}(x_j), x_j \in \mathrm{PA}(x_i)\})$. $m_{j \to i}(x_j)$ is the message going from $x_j$ to $x_i$ and is a function of $x_j$ (or $x_j(t)$ in the case of time dependent variables). The message $m_{j \to i}$ typically depends on some parameters of the edge $j \to i$. $\mathrm{agg}$ is a function that aggregates messages such as summation or multiplication. These operations are parallelizable and can be computed efficiently in graphs containing hundreds of thousands of edges via GPU acceleration[31].

**Optimal Transport.** Optimal Transport can be used to find a matching between two cell populations. Let us consider a source and a target population, $\{x^i\}_i$ and $\{y^j\}_j$, containing $p$ and $q$ cells respectively. $x^i, y^j \in \mathbb{R}^n$ are the expression profiles of the cells. In its discrete version, the optimal transport problem is to find a transport matrix $T \in \mathbb{R}^{p \times q}$ that minimizes the total cost:

$$\arg \min_{T \in \mathbb{R}^{p \times q}} c(T) = \sum_{i,j} T_{i,j} c(x^i, y^j)$$

such that $T.\mathbb{1}_q = \frac{1}{p}\mathbb{1}_p$, $T^T.\mathbb{1}_p = \frac{1}{q}\mathbb{1}_q$ and $T_{i,j} \geq 0 \; \forall i, j$. The cost function $c$ is typically a distance in $\mathbb{R}^n$. In the case where $p = q$, there exists a solution to the problem which is a permutation matrix, matching each cell $x^i$ in the source population to a unique cell $y^j$ in the target population[66].

**Posterior distribution.** Training a machine learning model $f_\theta$ on some dataset $\mathcal{D}$ typically results in a point estimate of parameters that best explain the data, $\theta^* \in \arg \max_\theta P(\mathcal{D} | \theta)$. However, the set of optimal parameters may be large, *i.e.* many different parameters $\theta^*$ may explain the data equally well. Bayesian methods impose a prior $P(\theta)$ on parameters and aim to estimate the whole distribution of parameters that are consistent with the data, $P(\theta | \mathcal{D}) \propto P(\mathcal{D} | \theta)P(\theta)$, called the posterior distribution. Deep ensembles[38] consist in independently training $k$ versions of the same model on the dataset $\mathcal{D}$, yielding a set of optimal parameters $\{\theta^*_1, ..., \theta^*_k\}$ that we informally refer to as independent samples from the posterior distribution, even if deep ensembles do not explicitly impose a prior $P(\theta)$ on parameters.

## Background on transcription modelling

**Dynamical models of transcription from microarray data.** Transcriptional regulation has been modeled through bottom-up approaches, wherein the assumptions and functional forms have been chosen from a precise understanding of the chemical interactions and equilibria[67] that underpin transcription mechanisms, such as the binding of transcription factors onto regulatory regions. These models rely on differential equations and have traditionally been applied at the very small scale, capturing only a handful of genes, focused on a specific gene's mechanism, and based on low dimensional measurements[68]. They have been adapted[69] and scaled-up[24,70] to the context of microarray data. Microarray is a high-throughput method for measuring gene expression at the bulk level.

Many variations have been explored, including Gaussian processes[71,72], jump processes[73] and Granger causality[74], some of them incorporating prior knowledge[75]. Some methods rely on derivative functions that directly depend on several past observation points[76]. Recently methods have leveraged neural ODEs[70] for microarray data, allowing them to work in continuous time and relax previously made assumptions. For instance, in PHOENIX[70], the derivative function still depends directly on time, providing more flexibility to the model, but not consistent with the goal of discovering time independent rules.

**Single-cell simulators.** Simulators of single-cell data have been developed[77–80] through bottom-up approaches, often taking genetic interactions into account. They can operate at a large scale over thousands of genes, and often rely on biologically grounded assumptions, but their parameters cannot be learnt from data, or have simply been hand-tuned[80].

## Model

**Trajectory Extraction Methods.** FLeCS is trained on time series of individual cells. We developed several approaches to extract trajectories from single-cell datasets and obtain such time series. The different trajectory extraction methods are compared in Extended Data Fig. 2.

Quantile method: we apply diffusion pseudotime[81] to order cells along a developmental trajectory and then assign cells to clusters based on the quantiles to which their pseudotime value belongs. We then compute optimal transport (OT) matchings between cell populations in consecutive quantiles. The transport cost is the euclidean distance in either PCA space (dimension 50) or UMAP space.

K-NN method: we first compute a k-nearest neighbors (k-NN) graph between cells ($k = 30$), based on the euclidean distance in either PCA or UMAP space. Edges in the k-NN graph are weighted according to the euclidean distance. Given a source and a target cell population, we first adjust the size of the source population by upsampling or downsampling to the same number of cells as in the target population. We then compute an OT matching between the two populations. Given a pair of cells $(x, y)$ matched through OT, we compute the shortest path between $x$ and $y$ within the k-NN graph. This provides trajectories of various lengths. We then downsample trajectories to the desired number of time points.

**The FLeCS model.** FLeCS is a Causal Kinetic Model trained on series of single-cell expression profiles $[(G_1(t), ..., G_n(t)), t = t_0, ..., t_{max}]$. The model is initialized to the expression profile at time $t_0$, dynamics are integrated forward in time to compute estimates of the expression profiles at times $t_1, ..., t_{max}$. Mean square error is computed between predictions and observations at times $t_1, ..., t_{max}$, and gradients with respect to model parameters are computed using the adjoint method.

Dynamics are parametrized as follows:

$$\frac{d}{dt}G_i(t) := \sigma\left[\sum_{G_j \in PA(G_i)} \omega_{ji} G_j(t) + b_i\right] - \alpha_i G_i(t) \quad \forall i \in 1, ..., n$$

Where $\sigma$ is the sigmoid activation function, and $\omega_{ji}$ is a scalar parameter associated with edge $j \to i$ of the gene network corresponding to the regulation strength of $G_j$ on $G_i$. $b_i$ is a bias controlling the basal transcription rate of $G_i$ in the absence of any regulation. Summation over the parents of $G_i$ is computed efficiently in parallel across all genes $1, ..., n$ via message passing. $\alpha_i$ is the exponential decay constant controlling the decay rate of $G_i$.

**FLeCS with latent variables.** Latent variables can be added to the FLeCS model (Extended Data Fig. 1a,c). Such latent variables can help capture unobserved aspects of the cellular state at time $t$, which can be inferred from observations of the past trajectory of the cell at times $s \leq t$, and potentially improving predictions regarding the future trajectory of the cell at times $s \geq t$. The information captured by such latent variables is however hard to interpret.

Let $X(t) = (G_1(t), ..., G_n(t), L_1(t), ..., L_k(t))$ be the state of the system where $\{G_i\}_i$ are genes whose expression is observed, and $\{L_i\}_i$ unobserved latent variables. Connections between genes are based on the architecture of the gene network, and connections are added between every pair $(G_i, L_j)$ and $(L_i, L_j)$.

Dynamics are parametrized in a similar way as before:

$$\frac{d}{dt}X_i(t) := \sigma\left[\sum_{X_j \in PA(X_i)} \omega_{ji} X_j(t) + b_i\right] - \alpha_i X_i(t) \quad \forall i \in 1, ..., n+k.$$

At time $t_0$, latent variables are sampled from a prior distribution, $L_i(t_0) \sim \delta(0) \; \forall i$ in our experiments. During training, only the gene variables $\{G_i\}_i$ are taken into account within the loss function (mean square error between predicted and observed expressions $G_i(t)$ at times $t_1, ..., t_{max}$). Experiments are performed with $k = 50$ latent variables.

**FLeCS with protein proxy variables.** Protein proxy variables can be added to the FLeCS model (Extended Data Fig. 1b,d). Such variables act as mediators between a regulator gene's expression level, and its regulated genes' expression levels. Modeling these variables can help capture delayed responses, caused, for instance, by a slow export of mRNA from the nucleus to the cytoplasm where it is translated: a spike in a given gene's expression first needs to increase the concentration of the protein it codes for, before the effect propagates to regulated genes. We expect protein proxy variables to capture information related to the concentrations of proteins, however one should refrain from interpreting them as the actual protein levels, given that such information is not measured in single-cell datasets.

Let $X(t) = (G_1(t), ..., G_n(t), P_1(t), ..., P_n(t))$ be the state of the system where $\{G_i\}_i$ are genes whose expression is observed, and $\{P_i\}_i$ their associated protein proxy variables, which are not observed. Connections from proteins to genes are based on the architecture of the gene network, and each gene is connected to the protein it codes for through an edge $G_i \to P_i$. No other connections are allowed.

Dynamics are parametrized in a similar way as before:

$$\frac{d}{dt}X_i(t) := \sigma\left[\sum_{X_j \in PA(X_i)} \omega_{ji} X_j(t) + b_i\right] - \alpha_i X_i(t) \qquad \forall i \in 1, ..., 2n.$$

At time $t_0$, protein proxy variables are sampled conditional on the expression of their associated coding genes, $P_i(t_0) \sim \delta(G_i(t_0))$ $\forall i$ in our experiments. During training, only the gene variables $\{G_i\}_i$ are taken into account within the loss function.

**Small molecule interventional model.** Given structural information about the drug, represented as a vector $v \in \mathbb{R}^l$, two multi-layer perceptrons, $E_p: \mathbb{R}^l \to \mathbb{R}^n$ and $E_d: \mathbb{R}^l \to \mathbb{R}^n$, predict the effects of the drug on production rates and decay rates, respectively. In our experiments, we use Morgan Fingerprints with radius 2 and 1024 bits. $E_p$ and $E_d$ have a single hidden layer of dimension $n$.

**Uncertainty quantification.** Samples from the parameters' posterior distribution are obtained via deep ensembles [38]. In our experiments, we use deep ensembles of size 10, therefore training 10 models independently for each condition, each providing one sample from the posterior distribution.

Uncertainties over individual parameters are obtained as the standard deviations of the posterior's marginals, $U(\theta_i, \mathcal{D}) = \sigma(\theta_i | \mathcal{D})$. When two distinct conditions are involved, uncertainty is computed as $U(\theta_i, \mathcal{D}_1, \mathcal{D}_2) = \sqrt{\sigma(\theta_i | \mathcal{D}_1)^2 + \sigma(\theta_i | \mathcal{D}_2)^2}$, which is proportional to the denominator of the Welch's t-statistic (assuming an equal number of samples from both posteriors).

## Datasets and preprocessing

**Datasets.** Our myeloid differentiation experiments rely on the Paul15 single-cell dataset[35], which was obtained from myeloid progenitor cells in the bone marrow of mice, and comprises several batches: (1) unsorted cells, (2) cells sorted based on Flt3 and Csf1r receptor protein expression, corresponding to common myeloid progenitors (CMPs), (3) cells with a CRISPR knockout on the Cepba gene, and (4) cells with a CRISPR knockout on the Cepbe gene.

The scATAC-seq derived network was obtained through the CellOracle[15] library using the function `data.load_mouse_scATAC_atlas_base_GRN()`. This network was constructed by identifying transcription factor binding motifs within accessible regulatory regions of genes. An interaction was added to the network whenever a binding motif was detected in a gene's regulatory region.

Our perturb-seq experiments rely on the Replogle-Weissman single-cell dataset[82], generated by performing CRISPR interference (CRISPRi) on human K562 cells, a myelogenous leukemia-derived cell line. CRISPRi was used to repress the expression of genes across the whole genome. Each cell is associated with a specific single-gene knockdown perturbation.

Chemical perturbation experiments are performed on the Sciplex3 single-cell dataset[51], which was generated through multiplexed chemical screening on human A549 cells, a cell line derived from adenocarcinomic epithelial cells. While the original dataset includes three cell lines, and 188 different compounds, we focus specifically on A549 cells and the 12 compounds that offered the clearest perturbation signals (see filtering procedure). For each compound, we use only cells treated with the highest dose.

The gene networks used for the last two datasets – the myeloid leukemia GRN and adenocarcinoma GRN –, were derived from the FANTOM5 database[48], which captured promoter activities through Cap Analysis of Gene Expression (CAGE). Promoter regions were screened for binding motifs of 662 transcription factors to derive 32 tissue-specific GRNs[83]. In our experiments, the myeloid leukemia and adenocarcinoma GRNs were selected to match as closely as possible the contexts of the cell lines from which single-cell data was acquired.

**Preprocessing.** The same preprocessing steps are applied to all datasets. First, the gene network is integrated with the single-cell dataset, retaining only genes present in both. Next, the single-cell data is processed using Scanpy[84] as follows: starting with raw counts, counts are normalized per cell, the top 4,000 highly variable genes are selected using Cell Ranger[85], and the data is denoised using the MAGIC algorithm[33].

**Filtering procedure.** Filtering was applied to the Replogle-Weissman to include only conditions with a sufficient number of cells. Specifically, we retain all knockdowns associated with at least 600 cells, resulting in a total of 13 conditions, including the control condition, after filtering.

For the Sciplex3 dataset, we filtered out conditions that are too similar to the control condition. To achieve this, we trained a binary classification boosting tree for each condition, and evaluated its accuracy on a held out test set. Perturbations with fewer than 50 cells or a test accuracy below $0.7$ were excluded.

**Cell cycle phase annotation.** Cells in the Replogle-Weissman single-cell dataset were labelled with cell cycle phases using `scanpy.tl.score_genes_cell_cycle`, based on lists of marker genes for the S and G2M phases[86].

# Training and technical details

**Training procedure.** FLeCS models are trained using the Adam optimizer[87] for all experiments. Myeloid differentiation experiments (learning rate: 0.005, batch size: 8, training data: 2048 trajectories for each condition). Model is trained on a mixture of trajectories of length 10 and length 3. During training on a knockout condition, the edges outgoing from the perturbed gene are removed.

Perturb-seq experiments (learning rate: 0.0005, batch size 8, training data: 1000 trajectories for each condition). Model is trained on a mixture of trajectories of lengths 2, 3, 4 and 6. During training on a knockdown condition, the edges outgoing from the perturbed gene are removed.

Sciplex3 experiments (learning rate: 0.0005, batch size 8, training data: 1024 trajectories of length 2 per condition). To stabilize training, data is augmented similarly to the FLow-Matching framework: given a training trajectory $(x(0), x(1))$, two time points $t_a$ and $t_b$ are sampled in $[0, 1]$, $t_a < t_b$, and the model is trained on a modified trajectory going from $(1 - t_a)x(0) + t_a x(1) + \varepsilon$ to $(1 - t_b)x(0) + t_b x(1)$, where $\varepsilon$ is some Gaussian noise with standard deviation proportional to $(t_b - t_a)(1 - t_b + t_a)$.

**Gene network sparsity influences model capacity.** The two FANTOM5-derived GRNs are significantly sparser, containing 0.7% and 0.8% of all possible edges, compared to the scATAC-seq derived GRN, which contains 3.8% of all possible edges. The sparsity of the gene network influences the capacity of the FLeCS model, and an overly sparse network can lead to underfitting. Therefore, when utilizing FANTOM5-derived GRNs, we introduce random edges to achieve a sparsity level comparable to that of the scATAC-seq derived network employed in the myeloid differentiation experiments. The observed underfitting suggests that FANTOM5-derived GRNs lack critical edges necessary to capture the observed cell dynamics under FLeCS' parameterization. Future research should explore more informed strategies for reducing network sparsity when necessary.

**Bounded production rates stabilize training.** The sigmoid activation function ensures that production rates remain bounded, regardless of the values of the parameters $\omega$ and $b$. This constraint is crucial to stabilize training, as the forward and backward passes in FLeCS rely on adaptive ODE solvers. These solvers adjust the number of time points at which they evaluate the dynamics $\frac{d}{dt}G$ based on the smoothness of the trajectory. Unbounded or erratic dynamics can lead to a significant increase in the computational costs associated with integrating these dynamics. This issue may arise even during later stages of the training process if the learned parameters result in non-smooth dynamics.

**Dynamics projection into UMAP space.** For visualization purposes, the dynamics learned by FLeCS in gene space are projected into UMAP space using the function `pl.velocity_embedding_stream` from scVelo[88], which is obtained based on transition probabilities between cells in accordance to their velocity vectors.

**Extraction of the most affected subgraph.** Given a score on edges, typically the absolute difference between the means of the posterior marginals for that edge across two different conditions, and given a desired number of genes $k$ ($k = 33$ in our experiments), we extract the most affected subgraph as follows: edges are ranked according to the score in descending order, we start by selecting the top $\frac{k}{2}$ edges, and greedily select more edges until $k$ distinct genes appear as either a source or target in at least one of the selected edges. We then consider the subgraph over these $k$ genes, including all edges from the original network among these $k$ genes.

In order to detect edges that are significantly affected by a perturbation, we perform Welch's t-tests to compare the means of the two marginal distributions (under the control and perturbed conditions). Welch's t-test does not assume equal variances between the two distributions. Given the large number of tests performed – one for each interaction in the network –, we adopt a conservative significance threshold ( $\frac{1}{\#edges}$) to limit the number of false positives. Without such adjustment, a typical significance threshold of 0.05 could result in approximately 30,000 false positive edges (among ~600,000 edges), assuming no true effects are present.

## Evaluation

**Velocity confidence.** Velocity confidence quantifies the local consistency between the inferred dynamics of neighbouring cells. This provides a measure of robustness to noise in the observations, which is especially interesting in the highly noisy context of single-cell data. For each cell $x_i$, we consider its set of neighbouring cells $\mathcal{N}(x_i)$, as defined by a k-nearest neighbour (k-NN) graph between cells. Denoting $v(x)$ the dynamics (or velocity) of a given cell $x$, velocity confidence for cell $x_i$ is defined as:

$$vc(x_i) = \frac{1}{k} \sum_{x_j \in \mathcal{N}(x_i)} corr(v(x_j), v(x_i))$$

where $corr$ refers to the Pearson correlation. Velocity confidence was computed using the function `tl.velocity_confidence` from scVelo[88].

**Benchmark methods for cell dynamics inference.** We benchmark FLeCS, along with its latent variable and protein proxy variants, against several methods. scTour[22] is a representation learning-based method for the inference of cellular dynamics. It typically operates in a 10-dimensional latent space, but we also evaluate a version with a latent space whose dimension matches the number of genes (4000). Optimal Transport Conditional Flow Matching[37] (OT-CFM) is an optimal transport-based generative method that has been applied to cell dynamics inference. Additionally, we include a baseline derived from the trajectories extracted from the data: given an extracted trajectory of observed cells $(x_0, ..., x_k)$ we estimate the dynamics of cell $i$ as $v(x_i) = (x_{i+1} - x_i), \forall i \in [0,..., k-1]$.

**Overrepresentation analysis workflow.** For different methods that identify sets of affected genes, we evaluate the alignment of these gene sets with known biological processes. The organism of interest is selected (Mus musculus for myeloid experiments and Homo sapiens for Perturb-seq and Sciplex experiments) and statistical overrepresentation tests are run for all GO[47] biological processes (10,7150 biological processes for humans and 10,458 for mice). Given the identified set of affected genes, overrepresentation analysis determines whether the overlap between the genes associated with a biological process, and the affected genes is larger than what would be expected by chance. For our overrepresentation analysis, we used the 4000 highly variable genes from the dataset as the reference set and conducted Fisher's exact tests to assess significance.

GO biological processes are organized hierarchically. When the affected gene set aligns closely with a specific process, we expect to observe overrepresentation in other processes within the same branch of the hierarchy. We examined the cumulative proportion of p-values from these tests across all biological processes. A higher proportion of small p-values indicates a stronger alignment with a particular branch of the GO hierarchy.

**Benchmark methods for gene set overrepresentation analysis.** GENIE3[13] is a tree-based method for the inference of GRNs from expression data. Originally developed for microarray data, it has been used in the context of single-cell data, for instance within the SCENIC[89] pipeline. Each gene is regressed from all other genes using Random Forests (with 100 trees in our experiments). Regulation interaction weights are then obtained through a measure of input variable importance. Given two GENIE3 models, each trained on a distinct condition, the most affected subnetwork is extracted in the exact same way as before, resulting in a set of $k$ affected genes.

We also evaluate against a variant of GENIE3 that leverages the gene networks that are used by FLeCS: each gene is regressed from its set of parents within the gene network.

Finally, we evaluate against standard differential expression (DE) analysis, using the function `tl.rank_genes_groups` from Scanpy, and the $k$ genes with the largest absolute z-score are retrieved. If retrieving genes that are most affected by two knockouts, differential expression analysis is run separately for each knockout, and the $k$ genes with the highest geometric average of the two absolute z-scores are selected.

**Lower bound on achievable Earth Mover distance.** Earth Mover's distance (EMD) between two populations of cells is computed based on the Euclidean distance in PCA space. To provide an estimate of the lowest achievable performance, we compute a lower bound as follows: we split the target population into two random halves, and compute the EMD between these two halves. Confidence interval on the lower bound is obtained through different splits of the target population.

**Evaluation of trajectory extraction methods.** We studied various aspects of the performance of the two trajectory extraction methods, quantile and k-NN, on the myeloid differentiation dataset. We evaluate the distributions of velocity confidence (vc), velocity norms, and L-smoothness for extracted trajectories of different lengths.

Given an extracted trajectory $(x_0,..., x_k)$, we estimate the smoothness of the trajectory in cell $x_i$ through:

$$L(x_i) = ||\frac{v(x_{i-1})}{||v(x_{i-1})||} - \frac{v(x_i)}{||v(x_i)||}|| \quad \forall i \in [1,..., k-1].$$

Lower values correspond to smoother dynamics around cell $x_i$. Normalized velocities are used in order to capture specifically changes in the direction of velocity, the distribution of velocity norms being studied separately.

When evaluated in gene space, the performance of the two methods is rather similar, the quantile method slightly outperforming the k-NN method (Extended Data Fig. 2a). However, if the pseudotime quantiles are not evenly distributed across the two differentiation branches, the quantile method extracts trajectories that jump across differentiation branches (Extended Data Fig. 2b). This limitation of the quantile method is further confirmed by a much heavier tail in the distribution of L-smoothness (in UMAP space), corresponding to sudden changes in velocity direction, and a heavier tail for the distribution of velocity norms (in gene expression space), corresponding to sudden jumps to different parts of the gene expression space (Extended Data Fig. 2c).

Besides, we evaluate the impact of building the k-NN graph using the euclidean distance between cells in either UMAP space or 50-dimensional PCA space (Extended Data Fig. 2c). The choice has no impact on the smoothness of trajectories (in gene space), velocity confidence is slightly better for short trajectories when the k-NN graph is built in PCA space, but this effect disappears, or is even reversed for longer trajectories. The biggest difference is in terms of smoothness in UMAP space, computing the k-NN graph in UMAP space leading to much smoother trajectories.

Therefore in the rest of our experiments we employed the following methods: the k-NN method for myeloid differentiation experiments, with a k-NN graph based on the euclidean distance in UMAP space; the quantile method for other experiments, with OT performed in UMAP space.

# Data availability

Raw published data for the Paul15, Replogle-Weissman and SciPlex3 single-cell datasets are available from the Gene Expression Omnibus under accession codes GSE72857, GSE146194 and GSE139944, respectively. The scATAC-seq derived network was obtained through the CellOracle library available at [https://github.com/morris-lab/CellOracle](https://github.com/morris-lab/CellOracle). All FANTOM5 derived Gene Regulatory Networks were downloaded from [http://www2.unil.ch/cbg/regulatorycircuits/Network_compendium.zip](http://www2.unil.ch/cbg/regulatorycircuits/Network_compendium.zip). Scripts to process the data are available at [https://github.com/Bertinus/FLeCS](https://github.com/Bertinus/FLeCS).

# Code availability

The FLeCS library, as well as scripts to reproduce experiments and figures are available at [https://github.com/Bertinus/FLeCS](https://github.com/Bertinus/FLeCS).

# Author contributions

P.B. conceptualized the study. P.B. developed and implemented FLeCS. P.B and J.V. standardized the codebase. A.T.L. and W.W. provided expert advice on metrics and datasets. P.B. performed the experiments, analyzed results and generated figures. P.B. wrote the initial manuscript. All authors gave feedback on the manuscript. S.B., F.J.T. and Y.B. supervised the research.

# Supplementary material

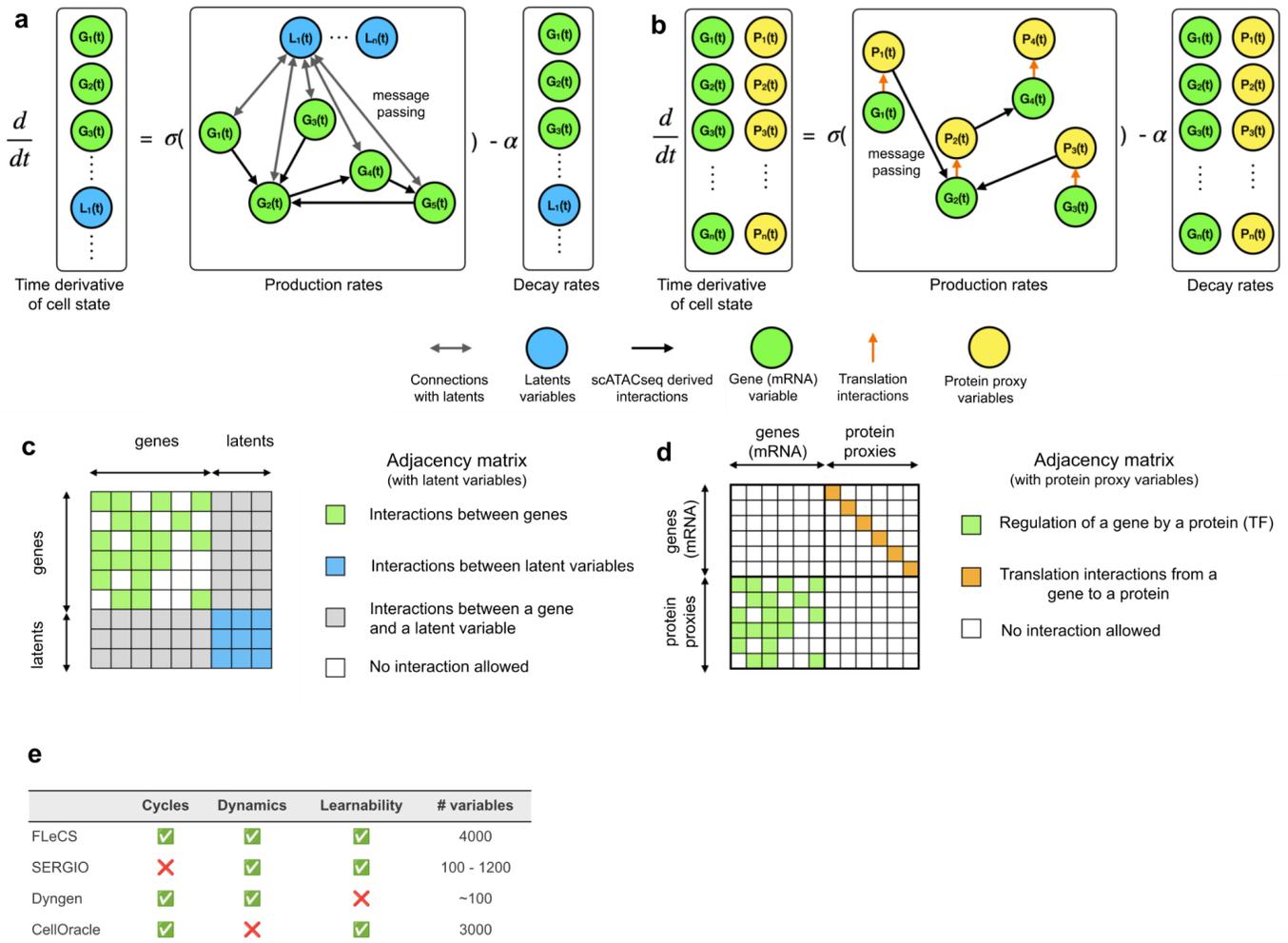

**Extended Data Fig. 1 | Overview of variants of FLeCS that model unobserved variables, and comparison with existing single-cell simulators. a**, The cell state can be extended to include time dependent latent variables, which are sampled from a fixed distribution at the initial time point. These latent variables take part in the computation of dynamics just like gene variables. **b**, The cell state can be extended to include protein proxy variables, which mediate the effect of a gene onto its regulated genes, and also depend on time. At the initial time point, a protein proxy variable is initialized with the same value as its coding gene variable. **c**, Adjacency matrix for FLeCS with latent variables. Dynamics involve gene-gene, gene-latent, latent-gene and latent-latent interactions, therefore mixing between traditional approaches (that rely on known variables only), and fully representation learning based approaches (that represent the whole system in latent space). **d**, Adjacency matrix for FLeCS with protein proxy variables. **e**, Comparison of FLeCS with existing single-cell simulators regarding their ability to model key aspects of transcriptional regulation, and scale.

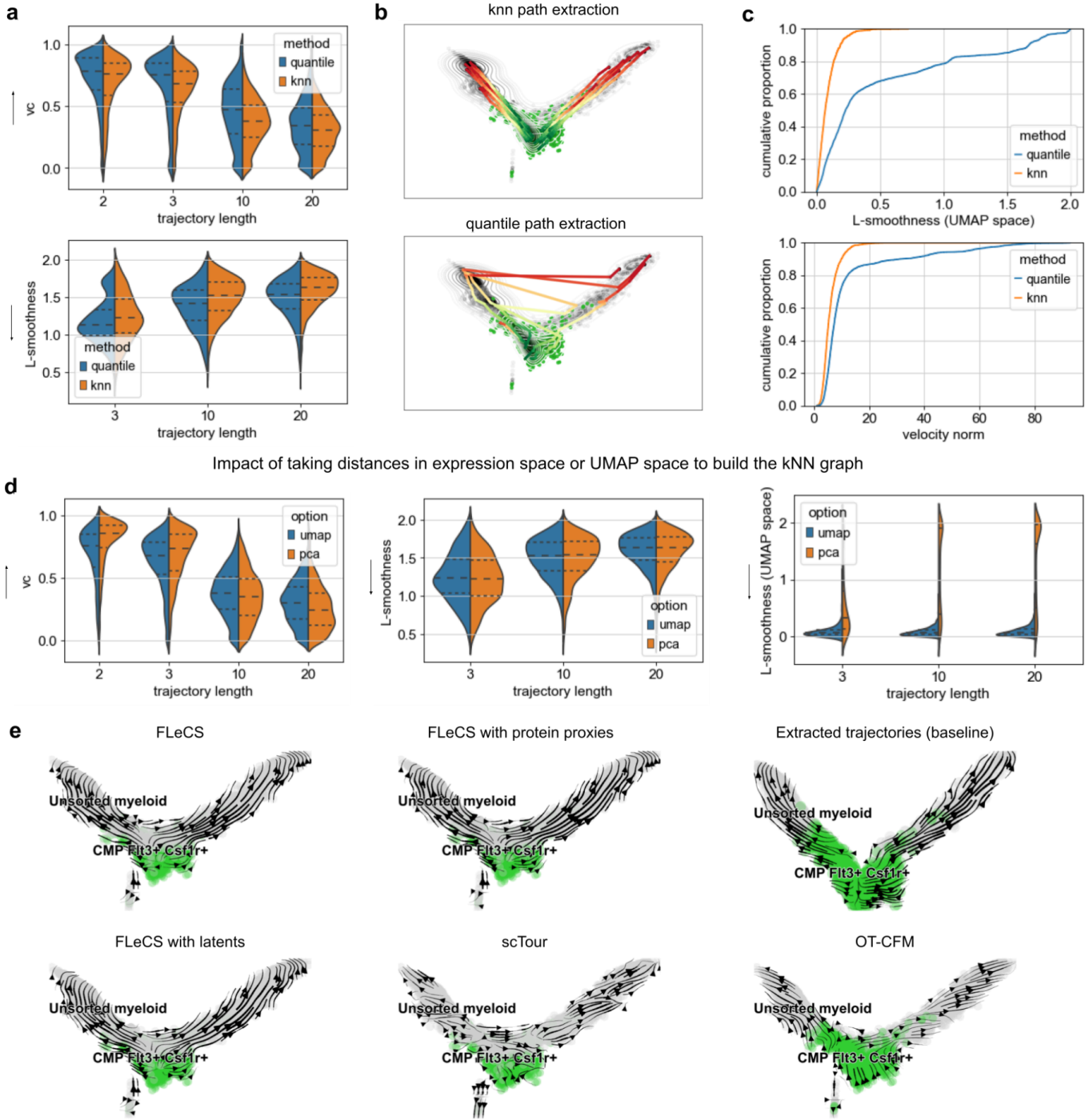

**Extended Data Fig. 2 | Comparison of trajectory extraction methods used during preprocessing, and visualization of learnt dynamics. a**, Distribution of velocity confidence (top) and L-smoothness (bottom) directly computed from extracted trajectories, in gene space. Results for trajectories of length 2 to 20 cells. The first extraction method (quantile) relies on OT matchings between consecutive quantiles of pseudotime. The second method (k-NN)

finds an OT matching between source and target cell populations, finds the shortest path between matched cells within the cell-cell k-NN graph, and subsamples the path to the desired length. **b**, Visualization of extracted trajectories using the knn method (top) and quantile method (bottom). **c**, Cumulative proportion of velocity length (top) and L-smoothness computed in UMAP space (bottom), for trajectories of 10 cells. **d**, Impact of taking distances in UMAP space or PCA space (dimension 50) to build the cell-cell k-NN graph, for the k-NN extraction method. Velocity confidence (left), L-smoothness in gene space (center) and L-smoothness in UMAP space (right). **e**, Projection into UMAP space of the dynamics learnt in gene space, for various methods.

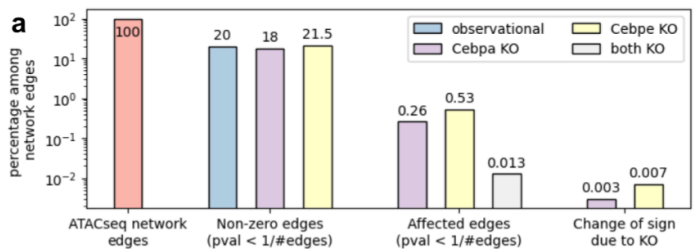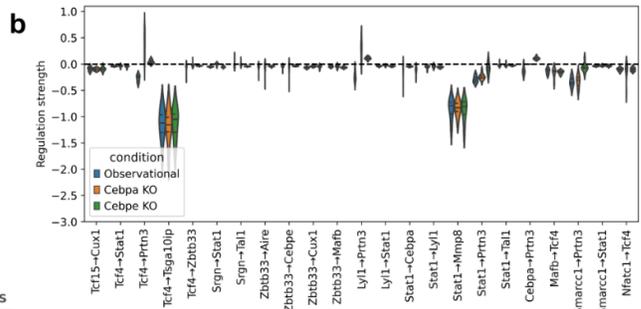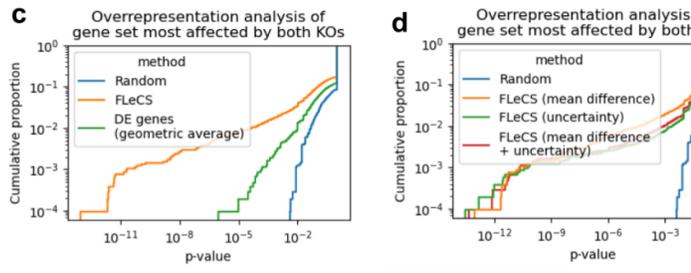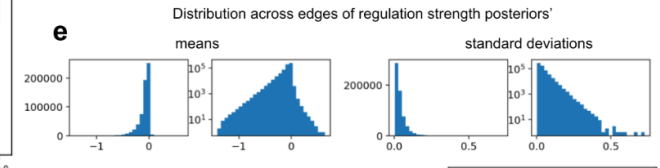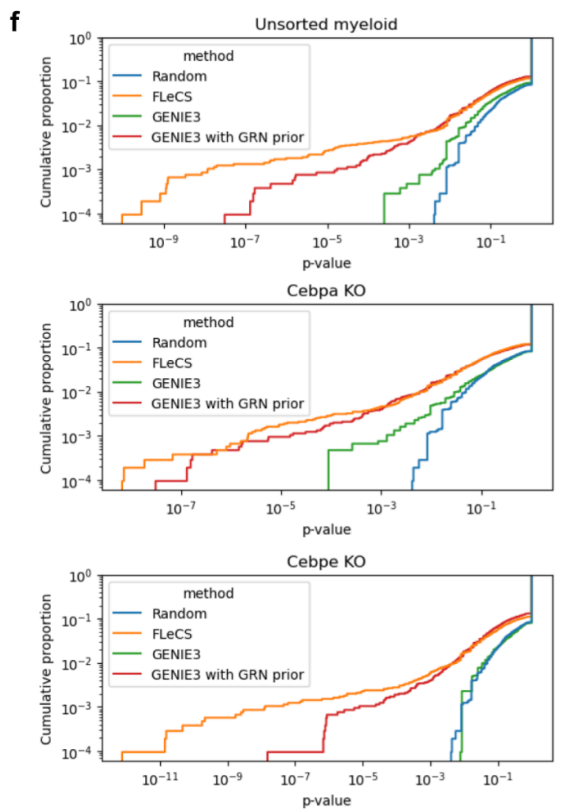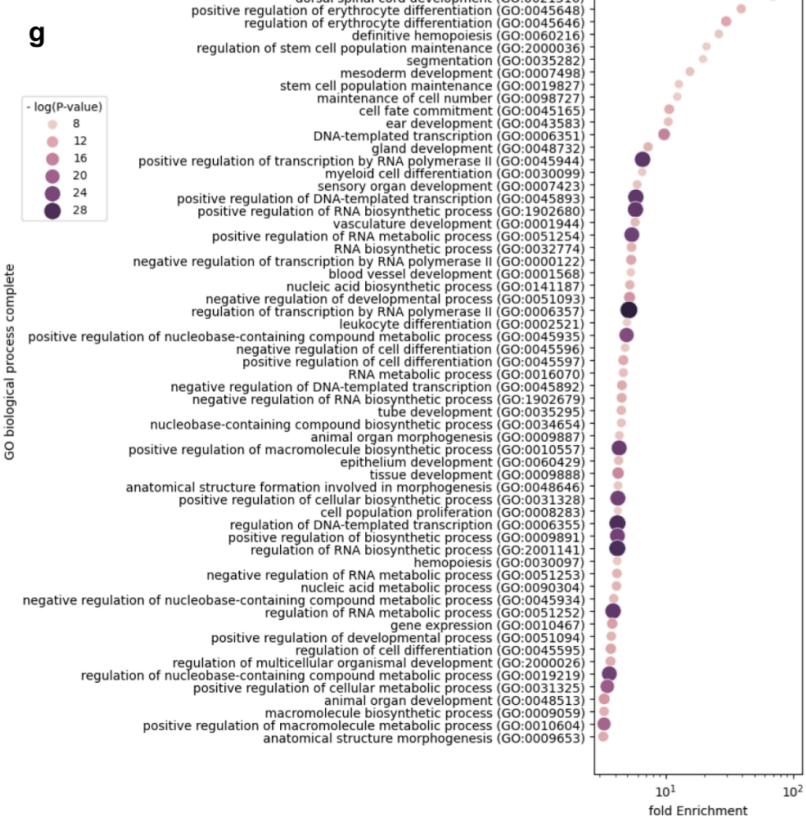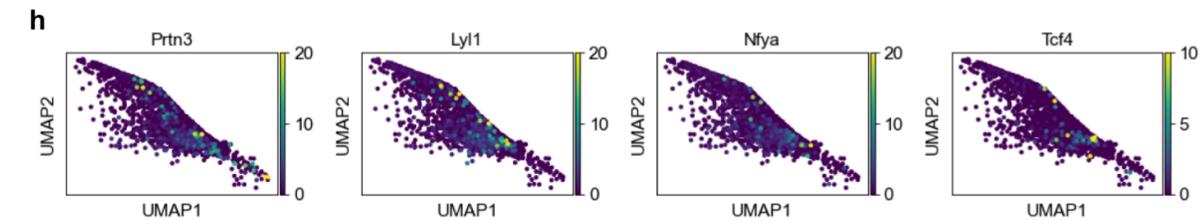

**Extended Data Fig. 3 | FLeCS identifies mechanisms perturbed by gene knockouts. a**, Histogram of various edge proportions among all edges in the scATAC-seq network. **b**, Regulation strength posteriors across the three conditions, for edges within the most affected subnetwork that are not shown in Fig. 3e. **c**, Overrepresentation of the gene set of the subnetwork most affected by both knockouts, across all biological processes in the PantherDB knowledgebase. Cumulative proportion of p-values from Fisher's exact tests, for FLeCS and the geometric average of differential expression p-values on the Cebpa and Cebpe knockouts. All sets have the same size. **d**, Comparison of different ways to select the most affected edges, based on the difference between the means of the posteriors, the uncertainty, or both. All sets have the same size. **e**, Distribution across edges of the means (left) and standard deviations (right) of the regulation strength posterior marginals. Distributions are both visualized on a linear and logarithmic scale. **f**, Overrepresentation of the gene set of the subnetwork consisting of strongest regulation interactions, across all biological processes in the PantherDB knowledgebase. All sets have the same size. **g**, Most enriched biological processes in the gene set of the subnetwork most affected by both knockouts. **h**, UMAP visualization of the megakaryocyte–erythroid lineage under the Cebpa knockout condition. Colors correspond to the expression levels of genes.

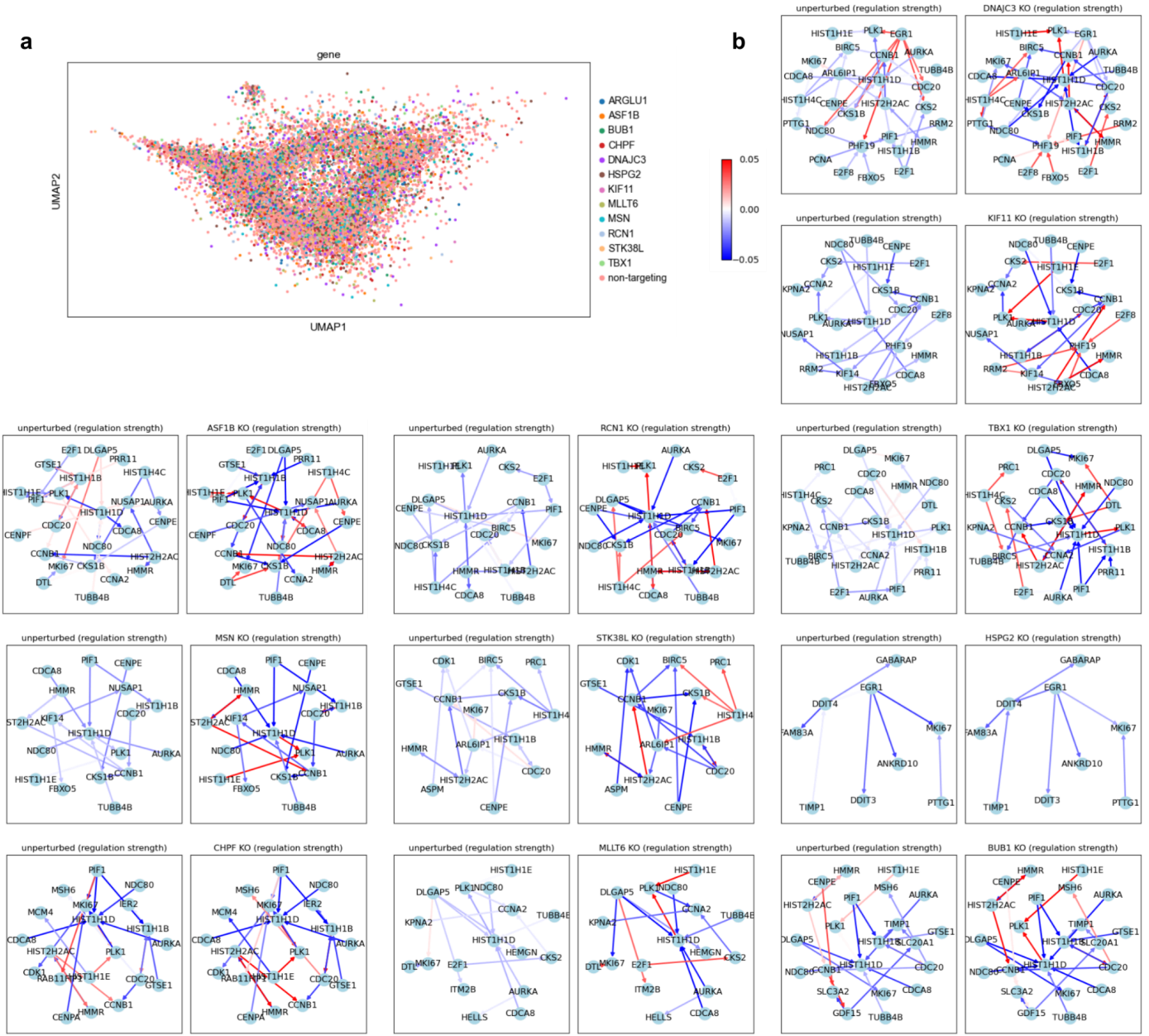

**Extended Data Fig. 4 | Knockdown cell populations, and most affected subnetworks identified by FLeCS from Perturb-seq data. a**, UMAP visualization of the different knockdown cell populations. **b**, Subnetwork most affected by the knockdown, for knockdowns that passed filtering. Edge colors represent the mean of the posterior, for unperturbed (left) and perturbed (right) cells.

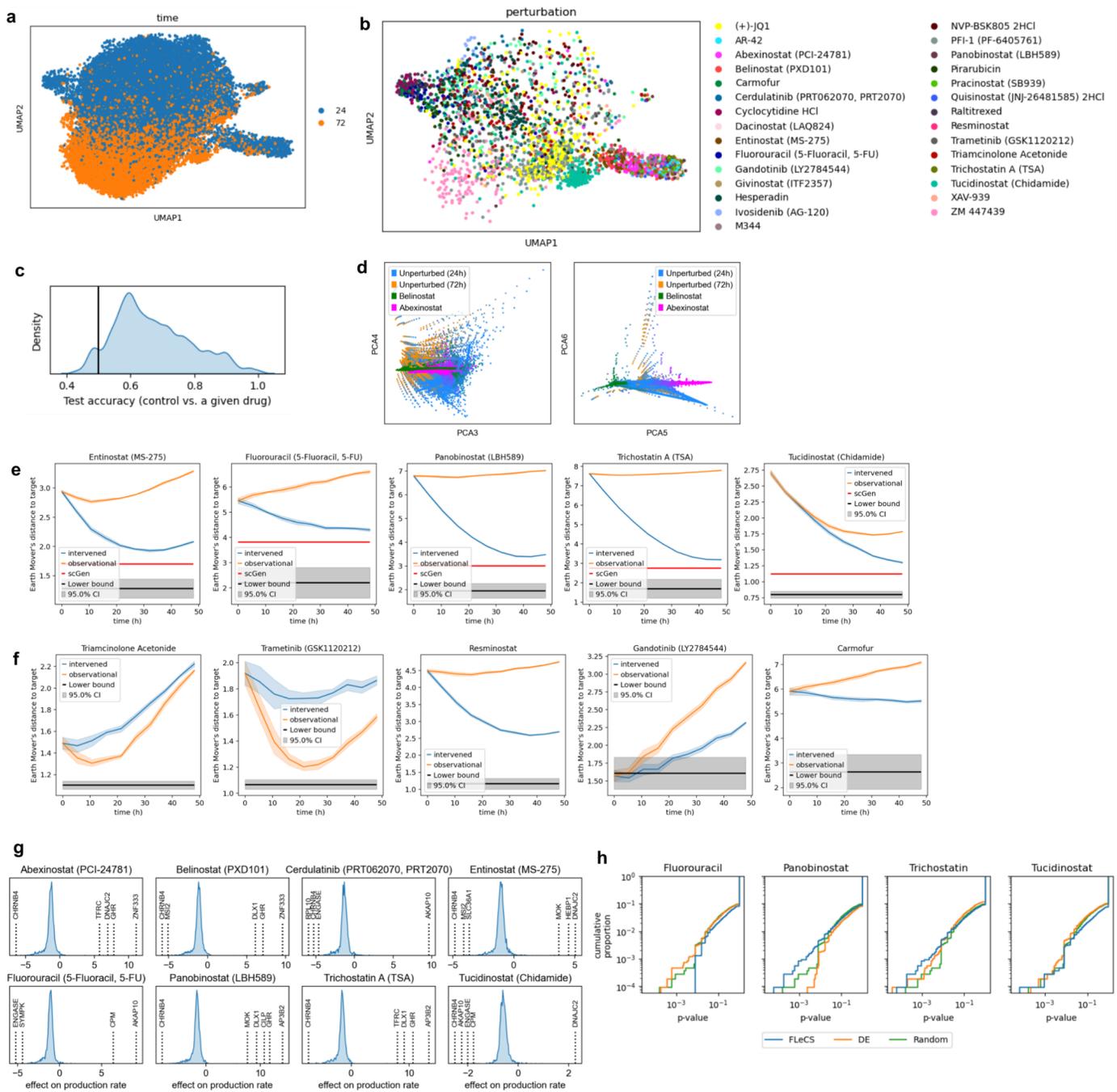

**Extended Data Fig. 5 | FLeCS captures small molecule perturbations. a**, UMAP representation of the drug screen dataset. Color corresponds to incubation time, 24h or 72h. **b**, UMAP visualization of the perturbed cell populations, for the different drugs. **c**, Distribution of test accuracies across all drugs. Simple binary classifiers are trained to distinguish between perturbed and control cells, for each drug. **d**, Principal component representation of synthetic cells generated by FLeCS for the control condition and under two drug perturbations, Abexinostat and Belinostat. **e**, Earth Mover's distance between generated cells and held out test cells from the target population, as a function of time along generated trajectories. Results for drugs included in the training set, comparing FLeCS in the control setting (observational), and when FLeCS is intervened on to capture drug effects. scGen does not take time into account,

therefore performance is represented as a horizontal line. A lower bound on achievable performance is shown in grey. **f**, Earth Mover's distance between generated cells and cells perturbed with novel drugs (on which FLeCS has not been trained). **g**, Distribution of drug effects on production rates across all genes, for 8 different drugs. Most affected genes are highlighted. **h**, Overrepresentation of the set of most affected genes across all biological processes in the PantherDB knowledgebase. Cumulative proportion of p-values from Fisher's exact tests. All sets have the same size.